\documentstyle[12pt]{article}
\newcommand{\journal}[4]{{\em #1~}#2\,(19#3)\,#4;}

\newcommand{\hpa}{\journal {Helv. Phys. Acta}}

\newcommand{\ijmp}{\journal {Int. J. Mod. Phys.}}

\newcommand{\pr}{\journal {Phys. Rev.}}

\newcommand{\cmp}{\journal {Comm. Math. Phys.}}
\newcommand{\cqg}{\journal {Class. Quantum Grav.}}

\newcommand{\np}{\journal {Nucl. Phys.}}
\newcommand{\pl}{\journal {Phys. Lett.}}

\newcommand{\prep}{\journal {Phys. Reports}}

\newcommand{\nc}{\journal {Nuovo Cim.}}

\makeatletter
\@addtoreset{equation}{section}
\makeatother
\renewcommand{\theequation}{\thesection.\arabic{equation}}
\newcommand{\dis}{\displaystyle}
\def\Lp{\displaystyle{\biggl(}}
\def\Rp{\displaystyle{\biggr)}}
\def\LP{\displaystyle{\Biggl(}}
\def\RP{\displaystyle{\Biggr)}}
\newcommand{\lp}{\left(}\newcommand{\rp}{\right)}
\newcommand{\lc}{\left[}\newcommand{\rc}{\right]}
\newcommand{\lac}{\left\{}\newcommand{\rac}{\right\}}
\newcommand{\G}{\Gamma}
\newcommand{\D}{\Delta}
\renewcommand{\a}{\alpha}
\renewcommand{\b}{\beta}
\renewcommand{\d}{\delta}
\newcommand{\e}{\varepsilon}
\newcommand{\eb}{\bar\varepsilon}
\newcommand{\f}{\phi}
\newcommand{\F}{\Phi}

\newcommand{\g}{\gamma}

\renewcommand{\l}{\lambda}
\newcommand{\lb}{\bar\lambda}

\newcommand{\m}{\mu}
\newcommand{\n}{\nu}

\newcommand{\mnrs}{\epsilon^{\mu\nu\rho\sigma}}
\renewcommand{\o}{\omega} 
\newcommand{\p}{\psi}
\renewcommand{\pb}{\bar\psi}
\newcommand{\r}{\rho}
\newcommand{\s}{\sigma} \renewcommand{\S}{\Sigma}

\newcommand{\z}{\zeta}

\newcommand{\BB}{{\cal B}}

\newcommand{\BS}{{\cal B}_\Sigma}

\newcommand{\FF}{{\cal F}}

\newcommand{\LL}{{\cal L}}

\newcommand{\NN}{{\cal N}}

\newcommand{\SS}{{\cal S}}

\newcommand{\WW}{{\cal W}}

\newcommand{\complex}{{\kern .1em {\raise .47ex
\hbox {$\scriptscriptstyle |$}}
    \kern -.4em {\rm C}}}
\newcommand{\real}{{{\rm I} \kern -.19em {\rm R}}}
\newcommand{\rational}{{\kern .1em {\raise .47ex
\hbox{$\scripscriptstyle |$}}
    \kern -.35em {\rm Q}}}
\renewcommand{\natural}{{\vrule height 1.6ex width
.05em depth 0ex \kern -.35em {\rm N}}}

\newcommand{\tr}{{\rm {Tr} \,}}
\newcommand{\cb}{{\bar c}}

\newcommand{\half}{\frac 1 2}
\newcommand{\pa}{\partial}
\newcommand{\pad}[2]{{\frac{\partial #1}{\partial #2}}}
\newcommand{\fud}[2]  {{\displaystyle{\frac{\delta #1}{\delta #2}}}}
\newcommand{\dfrac}[2]{{\displaystyle{\frac{#1}{#2}}}}

\newcommand{\etc}{{\em etc.\ }}

\newcommand{\sla}{\raise.15ex\hbox{$/$}\kern -.57em}

\newcommand{\twiddle}{\lower.9ex\rlap{$\kern -.1em\scriptstyle\sim$}}

\newcommand{\vf}{{\varphi}}
\newcommand{\pam}{{\partial_\mu}}
\newcommand{\pan}{{\partial_\nu}}

\renewcommand{\=}{=&} 
\newcommand{\equ}[1]{(\ref{#1})}

\newcommand{\eq}{\begin{equation}}
\newcommand{\eqn}[1]{\label{#1}\end{equation}}
\newcommand{\eea}{\end{eqnarray}}
\newcommand{\eqap}{\begin{eqnarray}}
\newcommand{\eqanp}[1]{\label{#1}\end{eqnarray}}
\newcommand{\eqa}{\eq\ba{rl}}
\newcommand{\eqan}[1]{\ea\label{#1}\end{equation}}
\newcommand{\ba}{\begin{array}}
\newcommand{\ea}{\end{array}}
\newcommand{\eqac}{\begin{equation}\begin{array}{rcl}}
\newcommand{\eqacn}[1]{\end{array}\label{#1}\end{equation}}

\renewcommand{\pad}[2]{{\displaystyle{\frac{\partial #1}{\partial #2}}}}

\def\intx{\displaystyle{\int d^4 \! x \, }}

\newcommand{\es}{\\[3mm]}

\newcommand{\fb}{\bar\f}

\newcommand{\Nb}{\pb^{*}{}}

\newcommand{\da}{{\dot{\a}}}
\newcommand{\db}{{\dot{\b}}}

\def\smab{\s^{\hspace{1mm}\m}_{\a\db}}
\def\esmab{\e^\a\smab\eb^{\db}}

\newcommand{\BSn}[1]{\BS^{(#1)}}
\def\TD{{\widetilde{\D}}}
\def\CD{{\D^{\#}}}
\def\HD{{\hat{\D}}}
\def\BD{{\bar{\D}}}

\newcommand{\reff}[1]{(\ref{#1})}

\newcommand{\ddsum}[2]{\dis{\sum_{#1}^{#2}}}

\def\Bmodd{\makebox[28mm]{$\BS$-modulo-$d$}}
\def\B0modd{\makebox[28mm]{$\BSn{0}$-modulo-$d$}}

\def\Pp{{P\, '}} 

\newcommand{\dpad}[2]{{\displaystyle \frac{\pa #1}{\pa #2}}}
\newcommand{\adot}{{\dot\alpha}}
\newcommand{\bdot}{{\dot\beta}}

\newcommand{\dsum}[1]{\dis{\sum_{#1}}}

\newcommand{\lhsb}{\dpad{}{\e^\a}\dfud{\G}{B^{*M}}}
\newcommand{\lhsf}{\dpad{}{\e^\a}\dfud{\G}{F_A}}
\newcommand{\caam}{c^A_{\a M}}
\newcommand{\dfud}[2]{{\displaystyle{\frac{\delta #1}{\delta #2}}}}

\begin{document}
\def\ftoday{{\sl  \number\day \space\ifcase\month
\or Janvier\or F\'evrier\or Mars\or avril\or Mai
\or Juin\or Juillet\or Ao\^ut\or Septembre\or Octobre
\or Novembre \or D\'ecembre\fi
\space  \number\year}}

\titlepage
\noindent
hep-th/9507045\\
UGVA-DPT-1995/07-897 \hfill July 1995


\begin{center}               \

{\huge Algebraic Renormalization of $N=1$ Supersymmetric
Gauge Theories\footnote{
Supported in part by the Swiss National Science Foundation.}}

\vspace{1cm}

{\Large Nicola Maggiore, Olivier Piguet and Sylvain Wolf}
\vspace{1cm}

{\it D\'epartement de Physique Th\'eorique --
     Universit\'e de Gen\`eve\\24, quai E. Ansermet -- CH-1211 Gen\`eve
     4\\Switzerland}

\end{center}
\vspace{15mm}

\begin{center}
\bf ABSTRACT
\end{center}

{\it The complete renormalization procedure of a
general $N=1$ supersymmetric gauge theory in the
Wess-Zumino gauge is presented, using
the regulator free ``algebraic renormalization'' procedure.
Both gauge invariance and supersymmetry are included into
one single BRS invariance.
The form of the general nonabelian anomaly is given.
Furthermore, it is explained how the
gauge BRS  and the supersymmetry
functional operators can be extracted from the general BRS operator.
It is then shown that the supersymmetry operators actually
belong to the closed,  finite,
Wess-Zumino  superalgebra when their action is restricted to
the space of the ``gauge invariant operators'', i.e. to the cohomology
classes of the gauge BRS operator.

 An erratum is added at the end of the paper.
}
\vfill

\newpage

\section{Introduction and Conclusions}

The first papers on supersymmetric gauge
theories and on their renormalization
appeared a long time ago~\cite{f-z-sym,s-s-sym,f-p-sym-ren}.
The renormalization of these theories
as well as the construction of gauge invariant
operators like, e.g., the supercurrent,
have been
completely performed, for the $N=1$ case, using the superspace
formalism~\cite{ps-book,p-s-ren}.

Theories with supersymmetry breaking remain to be
studied systematically.
The superspace renormalization schemes,
well adapted to
exact supersymmetry, may be extended to the case of
broken supersymmetry~\cite{ps-book}.
However, in such a situation, the main benefits of superspace
renormalization, such as manifest supersymmetry,
the nonrenormalization theorem of the
chiral interactions,
etc.~\cite{i-z-nonren,gates-book,ps-book}, are
lost in great part. It seems therefore desirable to be able
to master the renormalization procedure
 in terms of component fields which, in the Wess-Zumino gauge,
 presents the advantage of
using a much smaller number of field variables -- a
non negligible aspect
if one intends to perform explicit calculations.

In the Wess-Zumino gauge, however, the supersymmetry
transformations
act nonlinearly on the fields and, what is more critical, their
algebra is not closed~\cite{bm}.
Renormalization of supersymmetric gauge theories within this
framework
has yet been performed in the cases of  $N=4$~\cite{white2},
$N=2$~\cite{magg1,magg2}  and $N=1$~\cite{white1} supersymmetry.
The approach is based on a nilpotent generalized BRS operator
which combines
  the gauge invariance of the theory with
its rigid invariances -- supersymmetry in the present case.
The whole symmetry
algebra is translated into the nilpotency of
the generalized BRS operator. Generalized BRS invariance
is expressed, as usual, by a  Super Slavnov-Taylor (SST) identity. The
renormalizability proof and the search for the possible
anomalies are done using the method  of algebraic
renormalization~\cite{p-sor-book}, which has the effect
of reducing the analysis to a
cohomology problem in the space of the local field
polynomials.

The present paper deals with $N=1$ supersymmetric gauge theories,
without breaking, the case of broken supersymmetry being left for
subsequent publications~\cite{lmpw}. Beyond presenting the
formalism,  giving a simple proof of renormalizability and
deriving, as a new result, the explicit form of the gauge
anomaly, our aim is to answer the following question:
rigid symmetries -- supersymmetry in our case -- being
``hidden'' in the generalized BRS operator which also
contains gauge invariance, how the
rigid invariance -- or covariance -- of gauge invariant
operators can be characterized~?
Answering this question means finding a way to
extract the rigid symmetry generators from the BRS operator.
We shall see that this can be done, and
that it leads to an equivariant
cohomological structure~\cite{eq-cohom}. The nice outcome is that these
generators, when restricted to gauge invariant operators, i.e.
to cohomology classes of the {\it gauge} BRS operator, obey a closed
algebra. We conclude with the derivation of the Callan-Symanzik equation.

\section{The Model}

$N=1$ supersymmetry, involving fields with spin not greater than one,
is realized by means of two multiplets~\cite{ws}~:
\begin{description}
\item the matter multiplet $(\p^\a_a,\phi_a)$, consisting of a Weyl
fermion $\p^\a_a$ and a complex scalar field $\phi_a$, where $\a$ is a
spinorial index and the index $a$ runs over an arbitrary representation
of the gauge group, carried by the anti-Hermitian matrices $(T^i)_{ab}$;
\item the Yang-Mills multiplet $(A^i_\m,\l^{i\a})$, formed by the gauge
field and a Weyl fermion, both belonging to the adjoint representation
of the gauge group\footnote{Our conventions are listed in Appendix C.}.
\end{description}

In the Wess-Zumino gauge, the supersymmetry transformation laws are realized
nonlinearly, the nonlinearities being concentrated into the variations
of the spinors of the theory. As a consequence, the supersymmetry
algebra does not close simply on the translations, but it exhibits two kinds of
obstructions, represented by terms which vanish when equations of
motion are used (i.e. on-shell) and, in addition, by field dependent gauge
transformations~\cite{ws}~:
\eqa
[\d_1,\d_2]\F \= \!\!\! (\e^\a_1\smab\eb^\db_2 -\e^\a_2\smab\eb^\db_1)\pa_\m\F
\nonumber \\[3mm]
&\!\!\!
+\d_{\rm gauge}(\o^i)\F
\\[3mm]
&\!\!\!
+\ \mbox{field equations}\ ,\nonumber
\eqan{algebra}
where $\F$ collectively denotes all the fields of the theory, $\d$
describes the supersymmetry transformations having $\e^\a$ as infinitesimal
fermionic parameter, and $\d_{\rm gauge}(\o^i)\F$ stands
for gauge transformations with
field dependent parameter $\o^i$, which, for $N=1$ SYM is
\eq
\o^i\equiv(\e^\a_1\smab\eb^\db_2 -\e^\a_2\smab\eb^\db_1)A^i_\m.
\eqn{param}
Such an algebraic structure is a common feature of all supersymmetric gauge
field theories in the Wess-Zumino gauge,
and it has been shown in Ref.~\cite{bm}
that the usual approach consisting in treating separately gauge
invariance and supersymmetry leads
to a theory which requires infinitely many external sources to be renormalized,
because the presence of the field dependent gauge
transformations $\d_{\rm gauge}(\o^i)$ entails an hopelessly open algebra.
The introduction of auxiliary fields in order to
put the formalism off-shell, i.e. to eliminate from~\equ{algebra}
the presence of the equations of motion, even in the cases in which that is
possible, does not help to solve this problem.

An alternative and more convenient way to proceed is to collect all the
symmetries of the theory into a unique operator, and promoting the parameters
$\e^\a$ to the rank of global ghosts, with Faddeev-Popov $(\F\Pi)$ charge and
negative dimensions~\cite{dixon,white1,white2,magg1,magg2,mpr}.
The theory is consequently defined by a SST
identity, and the supersymmetry algebra is contained
into the simple nilpotency relation of the corresponding linearized
Slavnov-Taylor (ST) operator.

This same approach allowed to prove the perturbative finiteness of the
topological models~\cite{top},
which exhibit a supersymmetry-like algebraic structure,
and to study the renormalizability of Super Yang-Mills theories (SYM),
also in presence of extended supersymmetry~\cite{white2,magg1,magg2}.

The generalized BRS transformations, including ordinary BRS, supersymmetry,
translations and R-symmetry, are~:
\eqa
s A^{i}_\m \= \makebox[3mm]{}(D_\m c)^i
+\e^\a\s_{\m\a\db}\lb^{i\db} +\l^{i\a}\s_{\m\a\db}\eb^\db
           - i\xi^\n\pa_\n A^{i}_\m
\nonumber \es
s \l^{i\a}   \= -f^{ijk} c^j \l^{k\a}
                -\frac{1}{2}\e^\g\s^{\m\n}{}_\g{}^\a F^i_{\m\n}
             -\frac{i}{2}g^2\e^\a(\fb_aT^i_{ab}\f_b)
           - i\xi^\m\pa_\m \l^{i\a} -\eta\l^{i\a}
\nonumber \es
s\f_a \= -T^i_{ab}c^i\f_b +2\e^\a\p_{a\a}
            - i\xi^\m\pa_\m\f_a -\frac{2}{3}\eta\f_a
\nonumber\es
s\p^\a_a \= -T^i_{ab}c^i\p^\a_b -i\s^\m{}^\a{}_\db\eb^\db(D_\m\f)_a
             +2\e^\a \lb_{(abc)}\fb_b\fb_c
             -i\xi^\m\pa_\m\p^\a_a
             +\frac{1}{3}\eta\p^\a_a
\nonumber\es
sc^i \= -\frac{1}{2}f^{ijk}c^jc^k -2i\esmab A^i_\m -i\xi^\m\pa_\m c^i
\es
s\cb^i \= \makebox[3mm]{}b^i -i\xi^\m\pa_\m\cb^i
\nonumber\es
sb^i \= -2i \esmab\pa_\m\cb^i -i\xi^\m\pa_\m b^i
\nonumber\es
s\xi^\m \= -2\esmab
\nonumber\es
s\e^\a\=-\eta\e^\a
\nonumber\es
s\eta \=\makebox[3mm]{}0\ ,\nonumber
\eqan{brs}
where $c$, $\cb$ and $b$ are respectively the ghost, the antighost and the
Lagrange multiplier commonly introduced in order to fix the gauge,
and $\e^\a$, $\xi^\m$ and $\eta$ are the global
ghosts associated to the supersymmetry, the translations and the
R-symmetry\footnote{The global ghosts $\eta$ and $\xi$ are imaginary~:
\[
\bar\eta = - \eta,\quad \bar\xi = - \xi,
\]
and we recall the following conjugation rule~:
\[
\overline{s Bos} = s \overline{Bos},\quad \overline{s Fer} = - s
\overline{Fer},
\]
where $Bos$ and $Fer$ represent some bosonic or fermionic field, respectively.
}.
The covariant derivatives and the field strength are defined as follows~:
\begin{equation}
F^i_{\mu\nu} = \pa_\mu A^i_\nu - \pa_\nu A^i_\mu +  f^{ijk}A^j_\mu A^k_\nu\ ,
\end{equation}
\begin{equation}
\begin{array}{rcl}
(D_\mu \vf)^i &=& \partial_\mu \vf^i + f^{ijk}A^j_\mu\vf^k, \ \ \
\vf^i\in \hbox{adjoint representation} \\[3mm]
(D_\mu \psi)_a &=& \partial_\mu \psi_a + T^i_{ab}A^i_\mu  \psi_b, \ \ \
\p_a \in \hbox{matter field representation.}
\end{array}
\end{equation}

The most general action of dimension 4,
invariant under the transformation $s$, is
\eq
S=S_{SYM}+S_{gf},
\eqn{action}
where
\eqa
S_{SYM} = \intx \LP &\!\!\!
    \displaystyle{\frac{1}{g^2}}\lp
    -\frac{1}{4}F^{i\m\n}F^i_{\m\n}
    -i\l^{i\a}\smab(D_\m\lb^\db)^i \rp
    -\frac{1}{8}g^2 |\fb_aT^i_{ab}\f_b|^2
    +\frac{1}{2}|D_\m\f|^2
\nonumber\es
&\!\!\!
  -i\p^\a_a\smab(D_\m\pb^\db)_a
  -2\l_{(abc)}\f_b\f_c\lb_{(ade)}\fb_d\fb_e
  -i\pb_{a\db}T^i_{ab}\f_b\lb^{i\db }
\es
&\!\!\!
  -i\l^{i\a}\fb_aT^i_{ab}\p_{b\a}
  +2\l_{(abc)}\p^\a_a\p_{b\a}\f_c
  +2\lb_{(abc)}\pb_{a\db}\pb^\db_b\fb_c \RP,
     \nonumber
\eqan{sinv}
and
\eqa
S_{gf} \= s \intx \cb^i  \partial^\m A^i_\m  \es
\= \intx \LP b^i\partial^\m A^i_\m +  (\partial^\m\cb^i)\lp(D_\m c)^i
  +\e^\a\s_{\m\a\db}\lb^{i\db} +\l^{i\a}\s_{\m\a\db}\eb^\db\rp\RP.
\eqan{sgf}
The action $S$ depends on two kinds of coupling constants: the gauge coupling
constant $g$ and the Yukawa couplings $\l_{(abc)}$, which are
completely symmetric
in their indices. Notice that the fact
of dealing with a generalized BRS operator which contains all the symmetries
of the theory, allows us to solve easily, by means of the trivial
cocycle~\equ{sgf}, also another drawback affecting
the usual approach which keeps separate
the BRS transformations and the supersymmetry, i.e. the possibility
of defining an invariant gauge fixing term~\cite{bm}
(here we have adopted the Landau gauge).

The important property of the operator $s$ is to be nilpotent on-shell.
More precisely,~$s$~is exactly nilpotent on all the fields but the spinors,
on which its square gives equations of motion
\eq
s^2=\mbox{field equations}.
\eqn{nilponsh}
In order to get the on-shell nilpotency of the generalized BRS operator $s$,
it has been crucial that
the ghost $c$ transforms into the field dependent parameter
of the gauge transformations present in the algebra~\equ{algebra}, besides
translations and ordinary BRS variation.
We have thus successfully managed  to transform the algebraic
structure~\equ{algebra},
which is very difficult to handle, into a simple on-shell nilpotency relation.
Quantizing a theory defined by a on-shell nilpotent operator is a well
known problem~\cite{bv}.
To write the SST identity
corresponding to the generalized BRS transformations~\equ{brs},
it is sufficient
to add to the action a source term $S_{ext}$, which contains, besides the
usual external sources coupled to the nonlinear $s$-variations of the quantum
fields, also nonstandard terms, quadratic in the external sources.
This corrects for the fact that the $s$-operator is on-shell rather
than off-shell nilpotent~:
\eqa
S_{ext}=\intx\LP &\!\!\!
A^{*}{}^{i\m}(sA^i_\m) + \l^{*}{}^{i\a}(s\l^{i}_{\a})
+ \lb^{*}{}^i_\db(s\lb^{i\db})
+ \f^{*}_a(s\f_a) + \fb^{*}_a(s\fb_a) + \p^{*}{}^{\a}_{a}(s\p_{a\a})
\nonumber\es &\!\!\!
+ \Nb_{a\db}(s\pb^\db_a)
+c^{*i}(sc^i) -\frac{g^2}{2}(\e^\a \l^{*}{}^i_\a -\eb_\db\lb^{*}{}^{i\db})^2
+2\e^\a \p^{*}{}_{a\a}\eb_\db\Nb^\db_a \RP\ .
\eqan{sext}

We are now able to write, for the total classical action
\eq
\S=S_{SYM}+S_{gf}+S_{ext},
\eqn{classact}
the SST identity
\eq
\SS(\S)=0,
\eqn{sst1}
where
\eqa
\SS(\S)=\intx\LP &\!\!\!\!
\fud{\S}{A^{*}{}^{i\m}}
\fud{\S}{A^i_\m}
+
\fud{\S}{\l^{*}{}^i_\a}
\fud{\S}{\l^{i\a}}
+
\fud{\S}{\lb^{*}{}^{i\db}}
\fud{\S}{\lb^{i}_{\db}}
+
\fud{\S}{\f^{*}_a}
\fud{\S}{\f_a}
+
\fud{\S}{\fb^{*}_a}
\fud{\S}{\fb_a}
+
\fud{\S}{\p^{*}{}_{a \a}}
\fud{\S}{\p^\a_a}
\nonumber\\[3mm]&\!\!\!\!\!\!\!
+
\fud{\S}{\Nb_{a}^{\db}}
\fud{\S}{\pb_{a\db}}
+
\fud{\S}{c^{*i}}
\fud{\S}{c^i}
+
(b^i-i\xi^\m\pa_\m\cb^i)\fud{\S}{\cb^i}
\nonumber\\[3mm]&\!\!\!\!\!\!\!
+
\lp -2i\esmab\pa_\m\cb^i
-i\xi^\m\pa_\m b^i\rp\fud{\S}{b^i} \RP
-2\esmab\pad{\S}{\xi^\m}
-\eta\e^\a\pad{\S}{\e^\a}
-\eta\eb_\db\pad{\S}{\eb_\db}.
\eqan{slavnov}
For renormalization purposes, a relevant object is the linearized SST operator
\eqa
\BS =\intx\LP &\!\!\!\!
\fud{\S}{A^{*}{}^{i\m}}
\fud{}{A^i_\m}
+
\fud{\S}{A^i_\m}
\fud{}{A^{*}{}^{i\m}}
+
\fud{\S}{\l^{*}{}^i_\a}
\fud{}{\l^{i\a}}
-
\fud{\S}{\l^i_\a}
\fud{}{\l^{*}{}^{i\a}}
+
\fud{\S}{\lb^{*}{}^{i\db}}
\fud{}{\lb^{i}_{\db}}
-
\fud{\S}{\lb^{i\db}}
\fud{}{\lb^{*}{}^{i}_{\db}}
\nonumber\\[3mm] &\!\!\!\!
+
\fud{\S}{\f^{*}_a}
\fud{}{\f_a}
+
\fud{\S}{\f_a}
\fud{}{\f^{*}_a}
+
\fud{\S}{\fb^{*}_a}
\fud{}{\fb_a}
+
\fud{\S}{\fb_a}
\fud{}{\fb^{*}_a}
+
\fud{\S}{\p^{*}{}_{a \a}}
\fud{}{\p^\a_a}
-
\fud{\S}{\p_{a\a}}
\fud{}{\p^{*}{}^{\a}_a}
\nonumber\\[3mm] &\!\!\!\!
+
\fud{\S}{\Nb_{a}^{\db}}
\fud{}{\pb_{a\db}}
-
\fud{\S}{\pb_{a}^{\db}}
\fud{}{\Nb_{a\db}}
+
\fud{\S}{c^{*i}}
\fud{}{c^i}
+
\fud{\S}{c^i}
\fud{}{c^{*i}}
+
(b^i-i\xi^\m\pa_\m\cb^i)\fud{}{\cb^i}
\\[3mm]&\!\!\!\!
+
(-2i\esmab\pa_\m\cb^i -i\xi^\m\pa_\m b^i)\fud{}{b^i} \RP
-2\esmab\pad{}{\xi^\m}
-\eta\e^\a\pad{}{\e^\a}
-\eta\eb_\db\pad{}{\eb_\db}\ ,\nonumber
\eqan{slavnovlin}
which, by effect of~\equ{sst1}, is off-shell nilpotent
\eq
\BS^2=0.
\eqn{niloffsh}

The dimensions, Grassmann parities and R-weights of the fields are shown
in Table~\ref{table-dim}. The fields commute or anticommute according to the
formula:
\eq
\vf_1 \vf_2 = {(-1)}^{{\rm GP}_1 \cdot {\rm GP}_2} \vf_2 \vf_1 .
\end{equation}

\begin{table}[hbt]
\centering
\begin{tabular}{|c||c|c|c|c|c|c|c|c|c|c|c|c|c|c|c|}
\hline
&$A_\mu$&$\lambda$&$\f$&$\p$&$c$&$\cb$&$b$&$\xi^\m$
&$\e$&$\eta$&$A^*_\m$&$\l^*$&$\f^*$&$\p^*$ &$c^*$ \\
\hline\hline
$d$&1&3/2&1&3/2&0&2&2&-1&-1/2&0&3&5/2&3&5/2&4 \\
\hline
$GP$&0&1&0&1&1&1&0&1&0&1&1&0&1&0&0 \\
\hline
$\F\Pi$&0&0&0&0&1&-1&0&1&1&1&-1&-1&-1&-1&-2\\
\hline
$R$&0&-1&-2/3&1/3&0&0&0&0&-1&0&0&1&2/3&-1/3&0 \\
\hline
\end{tabular}
\caption[t1]{Dimensions   $d$, Grassmann parity $GP$, ghost number $\F\Pi$
  and R-weights.}
\label{table-dim}
\end{table}

\section{Renormalization}

According to the approach presented in the previous section,
the theory is formally characterized by the same
set of constraints defining the ordinary Yang-Mills theories~\cite{p-sor-book}.
In fact, $N=1$ SYM is defined by the following identities~:
\begin{description}
\item[the SST identity]
\eq
\SS(\S)=0,
\eqn{sst}
which contains the ordinary ST identity and the Ward identities for the
supersymmetry, for the translations and for the R-symmetry;
\item[the gauge condition]
\eq
\fud{\S}{b^i} = \pa^\m A^i_\m \ ;
\eqn{gaugecond}
\item[the ghost equation]
\eq
\FF^i\S=\D^i_{\rm g}\ ,
\eqn{ghosteq}
which is peculiar to the Landau gauge~\cite{bps}, with
\eq
\FF^i\equiv\intx\LP\fud{}{c^i}-f^{ijk}\cb^j\fud{}{b^k}\RP
\eqn{ghostexpr}
and where
\eqa
\D^i_{\rm g} \equiv \intx\LP &\!\!\!
f^{ijk}\Lp -A^{*}{}^{j\m}A^k_\m +\l^{*}{}^{j\a}\l^k_\a
+\lb^{*}{}^j_\db\lb^{k\db} +c^{*j}c^k \Rp
\nonumber\es &\!\!\!
+T^i_{ab} \Lp \f^{*}_a\f_b +\fb^{*}_a\fb_b
-\p^{*}{}^\a_ a\p_{b\a} -\Nb_{a\db}\pb^\db_b
\Rp\RP
\eqan{ghostbr}
is a classical breaking, i.e. is linear in the dynamical fields.
\item[the global ghost equations]
\eq
\pad{\S}{\xi^\m} = \D^{\rm t}_\m\ \ ,\ \ \pad{\S}{\eta} = \D_{\rm R}\ ,
\eqn{globalgh}
where $\D^{\rm t}_\m$ and $\D_{\rm R}$ are classical breakings given by
\eqa
\D^{\rm t}_\m \equiv -i\,\intx \LP&-A^{*}{}^i_\n\pam A^{i\n}
+\l^{*}{}^{i\a}\pam \l^i_\a
+\lb^{*}{}^i_\db\pam \lb^{i\db}+c^{*i}\pam c^i\nonumber\es
&-\f^{*}_a\pam\f_a-\fb^{*}_a\pam\fb_a+\p^{*}{}^\a_a\pam\p_{a\a}
+\Nb_{a\db}\pam\pb^\db_a\RP ,
\eqan{globalbr1}
\eq
\D_{\rm R} \equiv \intx \lp -\l^{*}{}^{i\a}\l^i_\a +\lb^{*}{}^i_\db\lb^{i\db}
+\frac{2}{3}\f^{*}_a\f_a-\frac{2}{3}\fb^{*}_a\fb_a
+\frac{1}{3} \p^{*}{}^\a_a\p_{a\a}-\frac{1}{3}\Nb_{a\db}\pb^\db_a
\rp .
\eqn{globalbr2}
\end{description}

We can then write the following algebra, valid for any functional $\g$
with zero GP~:
\eq
\BB_\g \SS(\g)  = 0 ,
\eqn{alg0}
\eq
\fud{}{b^i}\SS(\g)- \BB_\g\lp\fud{\g}{b^i}-\pam A^{i\m}\rp=\bar{\FF^i} \g  ,
\end{equation}
\eq
{\bar{\FF}}^i\SS(\g) +  \BB_\g {\bar{\FF}}^i \g = 0 ,
\end{equation}
\eq
\FF^i \SS(\g)+\BB_\g\lp\FF^i \g - \D^i_{\rm g}\rp = \WW^i_{rig}\g ,
\end{equation}
\eq
\WW^i_{rig}\SS(\g) - \BB_\g\WW^i_{rig}\g = 0 ,
\end{equation}
\eq
\pad{}{\xi^\m} \SS(\g)+ \BB_\g \lp\pad{\g}{\xi^\m}-\D^{\rm t}_\m\rp=\WW_\m \g
,
\end{equation}
\eq
\WW_\m \SS(\g) - \BB_\g \WW_\m \g = 0 ,
\end{equation}
\eq
\pad{}{\eta} \SS(\g)+\BB_\g \lp\pad{\g}{\eta}-\D_{\rm R}\rp = \WW_{\rm R} \g ,
\end{equation}
\eq
\WW_{\rm R} \SS(\g) -  \BB_\g \WW_{\rm R} \g = 0  ,
\eqn{alg8}
which also contains an antighost operator $\bar{\FF^i}$
and Ward operators for the rigid
transformations $\WW^i_{rig}$, for the translations $\WW_\m$ and for the
R-transformations $\WW_{\rm R}$, given by~:
\eq
{\bar{\FF}}^i \equiv \fud{}{\cb^i} + \pam \fud{}{A^{*}{}^i_\m} +
i\xi^\m\pa_\m \fud{}{b^i} ,
\eqn{antighost}
\eqa
\WW^i_{rig} \equiv \intx\LP&\!\!\!
-f^{jik}\lp
A^{k\m}\fud{}{A^{j\m}}
+A^{*}{}^{k\m}\fud{}{A^{*}{}^{j\m}}
+\l^{k\a}\fud{}{\l^{j\a}}
+\l^{*}{}^{k\a}\fud{}{\l^{*}{}^{j\a}}
-\lb^k_\db\fud{}{\lb^j_\db}
\right.\nonumber\\[3mm] &\!\!\!\left.
-\lb^{*}{}^k_\db\fud{}{\lb^{*}{}^j_\db}
+c^k\fud{}{c^j}
+c^{*}{}^k\fud{}{c^{*}{}^j}
+\cb^k\fud{}{\cb^j}
+b^k\fud{}{b^j}\rp
\nonumber\\[3mm] &\!\!\!
-T^i_{ab}\lp
\f_b\fud{}{\f_a}
+\f^{*}{}_b\fud{}{\f^{*}{}_a}
+\fb_b\fud{}{\fb_a}
+\fb^{*}{}_b\fud{}{\fb^{*}{}_a}
+\p^\a_b\fud{}{\p^\a_a}
+\p^{*}{}^\a_b\fud{}{\p^{*}{}^\a_a}
\right.\nonumber\\[3mm] &\!\!\!\left.
-\pb_{b\db}\fud{}{\pb_{a\db}}
-\pb^{*}{}_{b\db}\fud{}{\pb^{*}{}_{a\db}}\rp\RP
,
\eqan{wrig}
\eqa
\WW_\m \equiv -i \intx & \lp\dsum{\mbox{\scriptsize{all fields}}\; \vf}
\pam\vf\dfud{}{\vf}\rp ,
\eqan{wmu}
\eqa
\WW_{\rm R} \equiv\intx\LP &\!\!\!
-
\l^{i\a}\
\fud{}{\l^{i\a}}
+
\l^{*}{}^{i\a}
\fud{}{\l^{*}{}^{i\a}}
+
\lb^{i\db}
\fud{}{\lb^{i\db}}
-
\lb^{*}{}^{i\db}
\fud{}{\lb^{*}{}^{i\db}}
-
\frac{2}{3}\f_a
\fud{}{\f_a}
\nonumber\\[3mm] &\!\!\!
+
\frac{2}{3}\f^{*}_a
\fud{}{\f^{*}_a}
+
\frac{2}{3}\fb_a
\fud{}{\fb_a}
-
\frac{2}{3}\fb^{*}_a
\fud{}{\fb^{*}_a}
+
\frac{1}{3}\p^\a_a
\fud{}{\p^\a_a}
-
\frac{1}{3}\p^{*}{}_{a}^{\a}
\fud{}{\p^{*}{}_{a}^{\a}}
\nonumber\\[3mm] &\!\!\!
-
\frac{1}{3}\pb^\db_a
\fud{}{\pb^\db_a}
+
\frac{1}{3}\Nb_{a}^{\db}
\fud{}{\Nb_{a}^{\db}}
\RP
-\e^\a\pad{}{\e^\a}
-\eb_\db\pad{}{\eb_\db}
 ,
\eqan{wr}
where ``all fields" includes all the fields listed in Table 1.

Setting $\g\equiv\S$ in the relations~\reff{alg0} to~\reff{alg8} and
using the conditions~\reff{sst} to~\reff{globalgh} satisfied by the
classical action $\S$, leads to~:
\eq
{\bar{\FF}}^i \S = 0 , \quad
\WW^i_{rig} \S = 0 , \quad
\WW_\m  \S = 0 , \quad
\WW_{\rm R} \S = 0 .
\eqn{classcond}

The equations~\reff{ghosteq} and~\reff{globalgh} express thus the linearity
of the rigid transformations, of the translations and  of the
R-transformations.

The aim of the renormalization is to construct a quantum extension of the
theory, described  by the 1PI generating functional
\eq
\G = \S + O(\hbar),
\end{equation}
obeying the conditions~\reff{sst} to~\reff{globalgh}. By consequence of the
algebra~\reff{alg0} to~\reff{alg8}, it will also obey the
equations~\reff{classcond}.

Supersymmetric gauge field theories are characterized by the lack
of a coherent regularization scheme under which all the symmetries
of the theory, i.e. BRS and supersymmetry, are preserved. This implies
the adoption of a renormalization procedure not relying on a particular
kind of
regularization. The algebraic procedure of renormalization, based
on the general grounds of power counting and locality, satisfies this
requirement~\cite{p-sor-book}. Following this method,
the discussion of the quantum extension of
the theory is organized according to
two independent parts~:
\begin{enumerate}
\item The study of the stability of the classical action
      under radiative corrections. This amounts to the search of the
      possible invariant counterterms and to check that they all correspond to
      a renormalization of the free parameters of the classical theory,
      i.e. of the coupling constants and of the field amplitudes.
\item The search for anomalies, i.e. the investigation whether the symmetries
      of the theory survive at the quantum level.
\end{enumerate}

\subsection{Stability}

In order to check that the classical action is stable under radiative
corrections, i.e. that these can be reabsorbed through a redefinition
of the fields and the parameters of the theory,
we perturb the classical action $\S$ by a local functional $\S_c$, having
the same quantum numbers as $\S$, namely canonical dimensions four and
vanishing
$\F\Pi$-charge~:
\eq
\S\longrightarrow\S'\equiv\S+\z\S_c,
\eqn{pert}
where $\z$ is an infinitesimal parameter.

We then require that the perturbed action~$\S'$ satisfies
the constraints~\reff{sst} to~\reff{globalgh}
defining the theory. This implies that the perturbation $\S_c$
\begin{itemize}
\item [1.]does not depend on $b,\xi,\eta$;
\item [2.]does depend on the ghost $c$ only if differentiated.
\end{itemize}
Furthermore, it follows from the SST identity~\reff{sst} and
the algebra~\reff{alg0} to~\reff{alg8}
that~$\S_c$
\begin{itemize}
\item [3.]obeys the condition ${\bar{\FF}}^i \S_c = 0$,
which implies that $\S_c $
depend on $\cb$ and $A^{*}{}^\m$ only through the combination
\eq
{\hat{A}}^{*}{}^{i\m}\equiv\pa^\m\cb^i+A^{*}{}^{i\m};
\eqn{comb}
\item [4.]obeys the conditions of invariance
\eq
\WW^i_{rig} \S_c = 0 , \quad
\WW_\m  \S_c = 0 , \quad
\WW_{\rm R} \S_c = 0 .
\end{equation}
\end{itemize}

Finally, at the first order in $\z$, the SST identity~\equ{sst} imposed to the
perturbed action~$\S'$, translates into the following condition on the
perturbation~:
\begin{itemize}
\item [5.]
\eq
\BS\S_c=0.
\eqn{slacond}
\end{itemize}
The equation~\equ{slacond}
constitutes a cohomology problem, due to the nilpotency of the linearized
SST operator $\BS$ (see~\equ{niloffsh}).
Its solution can always be written as the sum of a
trivial cocycle~$\BS\hat\S$,
corresponding to fields renormalizations, which are unphysical, and one
or more elements belonging to the cohomology of $\BS$, i.e. which cannot be
written as $\BS$-variations~:
\eq
\S_c=\S_{\rm ph} + \BS\hat\S\ .
\eqn{contr}
The strategy we applied to construct the explicit form of these terms
is explained in the Appendix B  and we give here the solution~:
\eq
\S_{\rm ph}=
Z_g\pad{\S}{g} + Z_{(abc)}\pad{\S}{\l_{(abc)}}
+ \bar{Z}_{(abc)}\pad{\S}{\lb_{(abc)}}
\eqn{sphys}
\eq
\hat\S= \intx\Lp
Z_A ({\hat A}^{*}{}^{i\m}A^i_\m -  \l^{*}{}^{i\a}\l^i_\a
      - \lb^{*}{}^i_\db\lb^{i\db})
+ {Z_\phi}_{ab} (\f^{*}_a\f_b - \p^{*}{}^\a_a\p_{b\a})
+ {Z_{\bar\phi}}_{ab} (\fb^{*}_a\fb_b - \Nb_{a\db}\pb^\db_b)\Rp,
\eqn{diman}
where $Z_g, Z_A$ are arbitrary constants and $Z_{(abc)}, \bar{Z}_{(abc)},
{Z_\phi}_{ab}, {Z_{\bar\phi}}_{ab}$ are invariant tensors.
The relations between the renormalizations constants, appearing in~$\hat\S$,
of fields belonging to a same supermultiplet, are the consequence
of the observation, stated in Appendix A, according
to which the counterterm cannot depend on terms containing $\e
\hat{A}^{*},\e\f^{*}$ and their complex conjugates.
Notice also that the cohomology part
$\S_{\rm ph}$ depends on parameters which correspond
to possible multiplicative renormalizations (and
hence nonvanishing beta functions)
of the gauge coupling constant
and of the Yukawa couplings. This is an algebraic result which just shows
that the radiative corrections can be reabsorbed and that no new terms appear
at the quantum level, which was our aim.

\noindent{\bf Remark:} {\em
This result is in agreement with the generic situation.
In certain cases, however, a
nonrenormalization theorem~\cite{i-z-nonren} states
that the Yukawa couplings remain unrenormalized. But this has been shown
only within the superspace formalism~\cite{gates-book}. And
even in this case, finite renormalizations may occur if there are
massless particles~\cite{jjw}. Note also that a class of $N=1$ SYM theories
does exist, where no coupling constant renormalization occurs
at all~\cite{lps}.
}

\subsection{Anomalies}

In this subsection we will deal with the problem of defining a quantum vertex
functional which satisfies the SST identity, or, in other words, which
preserves the symmetries defining the classical theory.
For what concerns Super Yang-Mills theories within the superspace
approach, it has been demonstrated  in~\cite{p-s-ren,ps-book},
that the only anomaly affecting $N=1$ SYM is the
supersymmetric extension of the
standard Adler-Bardeen anomaly, and its explicit form
has been given in~\cite{absup}. To our best
knowledge, the same result for $N=1$
SYM in the Wess-Zumino gauge has  been obtained
in~\cite{white1} for the abelian
case, the expression for the nonabelian supersymmetric gauge anomaly written
in the Wess-Zumino gauge being still lacking.

In fact, the aim of the renormalization is to show that it is possible
to define a quantum vertex functional
\eq
\G = \S + O(\hbar),
\end{equation}
such that, as for the classical theory
\eq
S(\G) = 0 ,
\eqn{slavgamma}
\eq
\fud{\G}{b^i} = \pam A^{i\m}, \quad
\FF^i\G = \D^i_{\rm g}, \quad
\pad{\G}{\xi^\m} = \D^{\rm t}_\m, \quad
\pad{\G}{\eta} = \D_{\rm R}.
\eqn{qqq}
Our strategy will thus be the
following: we begin by imposing~\reff{qqq} and try
to solve~\reff{slavgamma}. But, before doing
the latter, it will be convenient to
require the validity of the conditions
\eq
{\bar{\FF}}^i\G  = 0, \quad
\WW^i_{rig} \G = 0 , \quad
\WW_\m \G = 0 , \quad
\WW_{\rm R} \G = 0 ,
\eqn{trans}
which anyhow follow from~\reff{slavgamma} and~\reff{qqq} through the
algebra~\reff{alg0} to~\reff{alg8}. Actually this program will fail, namely
the SST~\reff{slavgamma} will
turn out to be anomalous~:
\eq
\SS(\G) = r \D_{\mbox{\tiny{SAB}}},
\eqn{slavqu}
where $\D_{\mbox{\tiny{SAB}}}$ is the anomaly to be derived in the following
(see~\reff{sab} for the result) and $r$ is
a well-known function of the parameters of
the theory of order $\hbar\;$~\cite{ps-book,zinjust},
which
however cannot be determined by the pure algebraic method used here.

On the one hand, the extension of
the classical conditions~\reff{gaugecond}, \reff{ghosteq} and~\reff{globalgh}
to their quantum counterparts~\reff{qqq} is trivial (see ~\cite{p-sor-book}),
and the extension of the classical rigid invariances~\reff{classcond}
to their quantum
counterparts~\reff{trans} has been proven in~\cite{symbreak,p-sor-book}.
The search of the breaking $\D_{\mbox{\tiny{SAB}}}$ of the
SST identity~\reff{slavqu}, on the other hand,
requires some care. The rest of this section will
be devoted to this end.

According to the quantum action principle~\cite{qap,p-sor-book},
the SST identity gets a
quantum breaking
\eq
\SS(\G) = \D \cdot \G ,
\eqn{anomal-slavnov}
which, at the lowest order in $\hbar$, is a local integrated functional with
canonical dimension four and Faddeev-Popov charge one
\eq
\D \cdot \G  = \D + O(\hbar\D).
\end{equation}

The algebra~\reff{alg0} to~\reff{alg8}, written for the functional~$\G$,
at the lowest non-vanishing order in~$\hbar$
implies the following consistency conditions on the breaking $\D$~:
\eq
\fud{\D}{b^i} =  0,\quad
\FF^i \D = 0,\quad
\pad{\D}{\xi^\m} = 0,\quad
\pad{\D}{\eta} = 0,
\eqn{fant}
\eq
\bar{\FF^i} \D = 0,\quad
\WW^i_{rig} \D = 0, \quad
\WW_\m \D = 0, \quad
\WW_{\rm R} \D = 0,
\eqn{wrcons}
\eq
\BS \D = 0.
\eqn{brsl}
Notice that the consistency
conditions~\reff{fant} to~\reff{brsl} formally coincide
with the relations determining the counterterm. The difference is that now
the solution must belong to the space of local functionals having Faddeev-Popov
charge
one instead of zero. Therefore, the first eight  conditions tell us that the
breaking $\D$ does not depend on $b^i$, $\xi^\m$ and $\eta$,
that the ghost $c^i$ must always be differentiated,
that ${\bar c}^i$
and $A^{*}{}^{i\m}$ appear  only in the combination
${\hat{A}}^{*}{}^{i\m}$~\reff{comb} and that
$\D$ is invariant under the rigid transformations, the translations and
the R-transformations.

The last consistency condition~\reff{brsl} is
often called the Wess-Zumino consistency
condition.   Like  the corresponding one in
the zero-$\F\Pi$ sector~\reff{slacond},
it constitutes a cohomology problem. Its
solution can always be written as the sum
of a trivial cocycle $\BS\HD$, which can be absorbed in $\G$
as a local counterterm
$-\HD$, and one or more elements belonging to the cohomology of $\BS$, i.e.
which
cannot be written as $\BS$-variations~:
\eq
\D = \BS\HD + \CD
\eqn{solco}

The method applied to construct the explicit form of the anomaly $\CD$ is
explained in Appendix B  and leads to
\eqa
\CD\equiv r \D_{\mbox{\tiny{SAB}}}= r \intx\LP &\!\!\!
\mnrs\lac d^{ijk}\lp\pam c^i -
2\l^{i\a}\s_{\m\a\db}\eb^\db-2\e^\a\s_{\m\a\db}\lb^{i\db}\rp
A^j_\n \partial_\r A^k_\s\right.\nonumber\es &\!\!\!
+\left.D^{ijmk}\lp\frac{1}{12}\pam c^i-\frac{1}{4}
\l^{i\a}\s_{\m\a\db}\eb^\db-\frac{1}{4}\e^\a\s_{\m\a\db}\lb^{i\db}\rp A^j_\n
A^m_\r A^k_\s\rac\nonumber\es &\!\!\!
-3d^{ijk}\lp\e^\a\l^i_\a\lb^j_\db\lb^{k\db}-
\eb_\db\lb^{i\db}\l^{j\a}\l^k_\a\rp\RP,
\eqan{sab}
where  $d^{ijk}$  is the totally
symmetric invariant tensor of rank three given by
\eq
d^{ijk}\equiv d^{(ijk)}=\frac{1}{2}\tr \lp\tau^i\lac\tau^j , \tau^k\rac\rp,
\eqn{dijk}
and $D^{ijmk}$ is an invariant tensor of rank four given by
\eq
D^{ijmk}\equiv d^{nij}f^{nmk}+d^{nik}f^{njm}+d^{nim}f^{nkj}.
\eqn{Dijkm}

\section{Symmetry Generators}

All the rigid symmetries of the theory (supersymmetry, translations and
$R$-inv\-ar\-iance) have been included,
together with the original gauge BRS invariance, into a single
ST identity, which we assume from now on to be
free of anomaly (see \equ{slavqu})~:
\eq
\SS(\G)=0\ ,
\eqn{a-slavnov}
{}From its construction, based on the nilpotency of the generalized
BRS operator $s$ \equ{brs}, this  identity also
incorporates the full
algebra formed by the various symmetries.

However, since the original algebra was not closed,
it not obvious how one can recover the individual
symmetry generators and their algebra from \equ{a-slavnov}.

  The purpose of the present section is to show
that it is in general possible to obtain functional operators
which generate the ordinary gauge BRS transformations and the rigid
symmetries. Moreover, the generator
of the gauge BRS transformations is nilpotent, it
commutes with those of the rigid symmetries,
and the latter obey an algebra
which closes
 when their action is
restricted to the space of gauge invariant operators, i.e. to the
cohomology space of the gauge BRS operator.

Such generators are useful in the construction of
(super)multiplets of
gauge invariant operators, e.g. in the construction of the
Ferrara-Zumino supercurrent~\cite{f-z-sc} in the
Wess-Zumino gauge~\cite{lmpw}.

Let us expand the linearized, nilpotent SST operator $\BB_\G$,
corresponding to the SST identity \equ{a-slavnov},
according to the filtration operator
\eq
\NN = \e^\a\pad{}{\e^\a} + {\bar\e}^\adot\pad{}{{\bar\e}^\adot}
   + \xi^\m\pad{}{\xi^\m}
  + \eta\pad{}{\eta} \ .
\eqn{a-filter}
This operator counts the degree in the global ghosts. The expansion reads
\eq
\BB_\G=\dsum{n\ge0}{} \BB_n\ ,\quad{\rm with}\quad \lc \NN,\BB_n\rc=n\BB_n\ .
\eqn{a-filtr-bb}
The nilpotency of $\BB_\G$,
\eq
\BB_\G^2= 0\ ,
\eqn{a-nilpotency}
which follows from the SST identity \equ{a-slavnov},
 implies, at the orders 0,1 and 2, the identities
\eq\ba{l}
{\BB_0}^2=0\ ,\es
\lac \BB_0 , \BB_1 \rac=0\ ,\es
{\BB_1}^2 + \lac\BB_0,\BB_2\rac=0\ .
\ea\eqn{a-filtr-nilp}
respectively,
where $\{\cdot,\cdot\}$ means the anticommutator.
We first remark that $\BB_0$, i.e. the operator $\BB_\G$ taken at vanishing
global ghosts, coincides  with the ``usual'' linearized ST operator,
i.e. the one corresponding to the gauge BRS operator. It is this
BRS operator which is used to define the physical, i.e. gauge invariant,
quantum operators, as follows.

\noindent{\bf Definition:} {\it
A gauge invariant quantum operator is defined
by the cohomology classes of the BRS operator $\BB_0$. More explicitly,
such a gauge invariant operator is defined by
\begin{itemize}
\item[1)] an insertion $Q\cdot\G$, i.e.
the generating functional of the 1PI Green functions with the composite
field $Q$ inserted, obeying the invariance condition
\eq
\BB_0 \lp Q\cdot\G\rp =0\ ,
\eqn{a-inv-cond}
\item[2)] the equivalence relation
\eq\ba{ll}
Q\cdot\G \sim Q'\cdot\G \quad &\mbox{ if and only if }\quad
  Q\cdot\G - Q'\cdot\G = \BB_0 (\hat Q\cdot\G)\es
  &\qquad\mbox{ for some insertion}\quad \hat Q\ .
\ea\eqn{a-equiv-rel}
\end{itemize}
}
The second identity \equ{a-filtr-nilp} , which expresses the
commutativity of the gauge transformations with $\BB_1$,
allows to define the action of
$\BB_1$ in the space of the gauge invariant operators defined above.
Indeed, if $Q\cdot\G$ is a representative of a $\BB_0$-cohomology
class,
then $\BB_1(Q\cdot\G)$ is still a representative of such a class.

The third identity \equ{a-filtr-nilp} means that the
restriction of the operator $\BB_1$ to the space of the gauge invariant
operators is nilpotent. Indeed, on any representative $Q$ of a
$\BB_0$-cohomology class, one has
\eq
{\BB_1}^2 (Q\cdot\G) = -\BB_0\BB_2 (Q\cdot\G) \sim  0\ .
\eqn{a-nilp-bb1}
One may call this property
{\it ``equivariant nilpotency''}~\cite{eq-cohom}.

Eq. \equ{a-nilp-bb1} is the desired result. In order to see this, let us
define the rigid symmetry generators $\WW$ by the expansion
\eq
\BB_1 = \e^\a \WW_\a + {\bar\e}^\adot {\bar\WW}_\adot + \xi^\m \WW_\m
  + \eta\WW_R
  -2 \e^\a\s^\m_{\a\adot}{\bar\e}^\adot\pad{}{\xi^\m}
  -  \eta\e^\a\pad{}{\e^\a}
  +  \eta{\bar\e}^\adot\pad{}{{\bar\e}^\adot}\ ,
\eqn{a-def-symm-gen}
where we have separated the terms expressing the transformation laws of
the global ghosts, from the rest which is linear in these ghosts.
Then it is obvious that the equivariant nilpotency of $\BB_1$ implies the
{\it ``equivariant algebra''}
\eq\ba{l}
\lac \WW_\a ,{\bar\WW}_\adot \rac \sim 2\s^\m_{\a\adot}\WW_\m\ ,\es
\lc \WW_R,\WW_\a \rc \sim \WW_\a\ ,\qquad
\lc \WW_R,\bar\WW_\adot \rc \sim - \bar\WW_\adot\ ,\es
(\mbox{other (anti)commutators } \sim 0)\ .
\ea\eqn{a-equiv-alg}

\noindent {\bf Remark}: {\it
This result reproduces at the quantum level the classical
supersymmetry algebra which obviously closes on the translations, when
restricted  to the space of the classical gauge invariant operators.
}

Since the effective calculations are always done on representatives of the
cohomology classes defining the gauge invariant operators,
it is worthwhile to write down explicitly the algebra of the symmetry
generators including the elements pertaining to $\BB_2$. The latter are
defined by
\eq
\BB_2 = \e^\a{\bar\e}^\adot X_{\a\adot} +  \half\e^\a\e^\b Y_{\a\b}
    +  \half{\bar\e}^\adot{\bar\e}^\bdot \bar Y_{\adot\bdot}\ .
\eqn{def-x-y}
(There is no term in $\eta$ and $\xi$ due to the third and fourth of
conditions \equ{qqq} expressing the linearity of the $R$-transformations
and of the translations.)
The second and third  of Eqs. \equ{a-filtr-nilp} then yield
\eq\ba{l}
\lac \BB_0,\WW_\a \rac = \lc \BB_0,\WW_R \rc =0\ ,\es
\lac \WW_\a ,{\bar\WW}_\adot \rac = 2\s^\m_{\a\adot}\WW_\m
     -\lac \BB_0,X_{\a\adot} \rac  \ ,\es
\lac \WW_\a ,\WW_\b \rac = -\lac \BB_0,Y_{\a\b} \rac
  \quad\mbox{and conjugate equation}\ ,\es
\lc \WW_R,\WW_\a \rc = \WW_\a\ ,\qquad
\lc \WW_R,\bar\WW_\adot \rc = - \bar\WW_\adot\ .
\ea\eqn{alg-w-x-y}
This algebra does not close, each order in the filtration
bringing new symmetry generators.

\section{Nonrenormalization Theorem for the \hfill\break
    Anomaly}\label{th. de nonren}

We have still to show that the anomaly coefficient $r$ in the
anomalous SST identity \equ{slavqu} is not renormalized
or, more precisely,
that it vanishes to all orders if its one-loop order is equal
to zero.

This theorem follows from the
observation that the present theory reduces to an ordinary gauge
theory if one sets the global ghosts $\xi^\m$, $\e^\a$
and $\eta$ to zero. Then it suffices to refer to the known
proofs~\cite{non-ren}.

\section{Callan-Symanzik Equation}\label{eq de callan-sym}
\renewcommand{\lb}{{\bar\l}}
\renewcommand{\pb}{{\bar\p}}
\renewcommand{\fb}{{\bar\f}}

Although the theory we are considering
is massless, a mass parameter $\m$
must be introduced in order to define the quantum theory. This
parameter fixes the scale where the normalization conditions
fixing the free parameters of the theory are taken.
Being the only dimensionful parameter in the present case, it
also determines the overall scale of the theory. This scale is
controlled by the Callan-Symanzik equation,
which follows from the renormalizability of the
model~\cite{p-sor-book}.

In order to derive the Callan-Symanzik equation, we first observe
that the scale operator $\m\,\pa/\pa\m$ ``commutes`'' with the
ST operator:
\eq
\m\dpad{}{\m}\SS(\G) = \BB_\G \lp\m\dpad{\G}{\m}\rp \ .
\eqn{dil-op}
The assumed validity of the SST identity
without anomaly and the quantum action principle thus imply
that the local, dimension 4, insertion  $\m\,\pa\G/\pa\m$
is $\BB_\G$-invariant. Let us expand the latter into a basis of
insertions with the same properties.
An appropriate basis is given by the following insertions, which
are the quantum extensions of the classical counterterms
\equ{sphys}, \equ{diman}:
\eq\ba{l}
\dpad{\G}{g}\ ,\quad \dpad{\G}{\l_{abc}} ,
\quad \dpad{\G}{\lb_{abc}},\\[4mm]
\NN_A\G=\BB_\G\intx\lp{\hat A}^*A-\l^*\l-\lb^*\lb  \rp \es
\phantom{\NN_A\G}
= \lp N_A+N_\l+N_\lb-N_{A^*}-N_{\l^*}-N_{\lb^*}-N_b-N_\cb \rp\G ,\es
\NN_{ab}\G = \BB_\G\intx\lp  \f_a\f^*_b-\p_a\p^*_b \rp =
  \lp N_\f+N_\p-N_{\f^*}-N_{\p^*}\rp_{ab}\G ,\es
{\bar\NN}_{ab}\G = \BB_\G\intx
        \lp \fb_a\fb^*_{b}-\pb_a\pb^*_b \rp =
  \lp N_\fb+N_\pb-N_{\fb^*}-N_{\pb^*}\rp_{ab}\G ,
\ea\eqn{inv-basis}
where we have introduced the counting operators
\eq\ba{l}
N_\vf = \intx \vf\dfud{}{\vf}\ ,\quad \vf =
 A,\l,\lb,A^*,\l^*,\lb^*,b,\cb\ ,\es
{N_\vf}_{ab} = \intx \vf_a\dfud{}{\vf_b}\ ,\quad
  {N_{\vf^*}}_{ab} = \intx \vf^*_b\dfud{}{\vf^*_a}\ ,\quad
 \vf=\f,\p,\fb,\pb\ .
\ea\eqn{count-op}
Expanding now the scale operator in this basis we obtain the
Callan-Symanzik equation
\eq
\lp \m\dpad{}{\mu} + \b_g\dpad{}{g} + \b_{abc}\dpad{}{\l_{abc}}
+ \bar\b_{abc}\dpad{}{\lb_{abc}}
 -\g_A\NN_A - \g_{ab}\NN_{ab}
   - {\bar\g}_{ab}{\bar\NN}_{ab}\rp \G = 0.
\eqn{Callan-Sym}
The $\b$-functions $\b_g$, $\b_{abc}$, as well as the anomalous
dimensions $\g_A$, $\g_{ab}$ are of order $\hbar$ and are
calculable in terms of vertex functions using the normalization
conditions.


\vskip5mm
\noindent {\bf Acknowledgments:}
We would like to thank Carlo Becchi and Claudio Lucchesi
for useful discussions.

\appendix
\renewcommand{\theequation}{\Alph{section}.\arabic{equation}}
\renewcommand{\thesection}{\Alph{section}}

\setcounter{equation}{0}
\setcounter{section}{1}
\section*{Appendix \thesection . Nonrenormalization of the Linear
         \break Supersymmetry Transformations}

The linearity of the supersymmetric transformations of the
bosonic fields $A_\m$, $\f$ is expressed, at the classical level,
by the fact that the right-hand sides of the equations
\eq\ba{l}
\dpad{}{\e^\a}\dfud{\S}{{\hat A}_\m^{*i}} =
   \s^\m_{\a\adot} \lb^{i\adot},\quad
\dpad{}{\e^\a}\dfud{\S}{\lb^i_\adot}
= -{\hat A}_\m^{*i}\s^{\m\adot}_\a ,\es
\dpad{}{\e^\a}\dfud{\S}{\f^*_a} = 2\p_{a\a},\quad
\dpad{}{\e^\a}\dfud{\S}{\p_{a\b}} = 2\f^*_a\d^\b_\a,
\ea\eqn{a1}
are at most linear in the dynamical fields.

In the simplifying notation
\[
\{B^{*M}\} = \{ {\hat A}^{*i}_\m,
\ \f^{*}_a\},\quad \{F^A\} = \{\lb^{i_\da},\ \p^\a_a\},
\]
(and similarly for the associated fields $B_M$,
$F^*_A$),  this reads
\eq
\dpad{}{\e^\a}\dfud{\S}{B^{*M}} = \caam F_A,\quad
\dpad{}{\e^\a}\dfud{\S}{F_A} = \caam B^{*M} .
\eqn{a2}

We want to show that, at all orders of perturbation theory:
\eq\ba{l}
\lhsb = \caam F_A + \D_\a\cdot\D_M\cdot\G = \caam F_A +
                                                  O(F^*),\es
\lhsf = \caam B^{*M} + \D'_\a\cdot\D'{}^A\cdot\G =
            \caam B^{*M} + O(F^*F) + O(F^*F^*) ,
\ea\eqn{a3}
i.e. that the classical results hold up to terms vanishing with
$F^*$. This implies in particular that the vertices
$\e^\a B^{*M}\caam F_A$, i.e.
\eq
{\hat A}_\m^{*i}\e\s^\m\lb^i,\quad
         {\hat A}_\m^{*i}\l^i\s^\m\bar\e,\quad
2\f^*_a\e\p_a, \quad 2\fb^*_a\eb\pb_a,
\eqn{aa4}
{\it are not renormalized}.

\noindent{\bf Remark:} {\em
All the graphs contributing to the radiative corrections to
\equ{a1} or \equ{a2} are superficially convergent.}

The proof of \equ{a3} goes by induction. Let us assume it to
hold at order $\hbar^{n-1}$:
\eq\ba{l}
\lhsb = \caam F_A +O(F^*) + O(\hbar^n),\es
\lhsf = \caam B^{*M} + O(F^*F)+O(F^*F^*) +O(\hbar^n).
\ea\eqn{aa5}
The quantum action principle together with the induction
hypothesis imply (for $n\ge1$)
\eq\ba{l}
\lhsb = \caam F_A + \D_\a\cdot\D_M\cdot\G
        +\hbar^n\D_{\a M} + O(\hbar^{n+1}),\es
\lhsf = \caam +B^{*M} + \D'_\a\cdot\D'{}^A\cdot\G
        +\hbar^n\D'_\a{}^A  + O(\hbar^{n+1}).
\ea\eqn{a6}
Let us begin with the first equation. The double insertion
$\D_\a\cdot\D_M\cdot\G$
corresponds to Feynman graphs where the derivatives with respect
to $\e$ and $B^*$ act on two different vertices, whereas the
single, local, insertion $\D_{\a M}$ corresponds to both
derivative
acting on a same vertex. For the latter we have made explicit
that
only the tree graphs contribute at its lowest non vanishing order --
order $\hbar^n$ by the induction hypothesis.

The double insertion graphs containing at least one loop, $\D_\a$
and $\D_M$ are produced by terms of the action of order
$n-1$ at most. Furthermore, since one of them must contain a
factor $\e$ and since, by hypothesis, $\e$ couples to $B^*$
only
in a trivial manner -- i.e. linearly in the dynamical fields --
 up to this order, it is its coupling with
$F^*$ which is involved. Thus
\eq
\D_\a\cdot\D_M\cdot\G = O(F^*).
\eqn{a7}
Similarly, for the second of the equations \equ{a6}, one obtains
\eq
\D'_\a\cdot\D'{}^A\cdot\G = O(F^*F) + O(F^*F^*).
\eqn{a8}
The additional dependence on $F$ or $F^*$ is implied by the
conservation of the total number of these spinor fields, which
can easily be checked by inspection of the action.

Let us come now to the single insertions $\D_{\a M}$ and
$\D'_\a{}^A$. They are field polynomials of dimensions bounded
by the dimensions of the left-hand sides of \equ{a6}, i.e.
by 3/2 and~3, respectively. Moreover their ghost numbers and
$R$-weights are those of the left-hand sides, too. A detailed
analysis then shows that they have exactly the same form as the
right-hand sides of \equ{a1}. In our notation:
\eq
\D_{\a M} = r^A_{\a M} F_A,\quad \D'_\a{}^A = {r'}_{\a M}^A
                                                     B^{*M}.
\eqn{a9}
The coefficients $r$ and $r'$ are evaluated by computing the
3-point vertex
\eq
\G_{\e^\a B^{*M} F_A} =
  \left.\dpad{}{\e^\a}\dfud{}{B^{*M}}\dfud{}{F_A}
    \G\right\vert_{{\rm all\ fields}\ \vf=0}
\eqn{a10}
at the order $\hbar^n$ ($n\ge1$). But, since the coupling of
$\e$ with
$B^*$ is linear in $F$ up to the order $n-1$ by the induction
hypothesis, there is no one-particle-irreducible loop graph
contributing to \equ{a10}.
Thus the coefficients $r$ and $r'$ vanish.
This ends the proof of \equ{a3}.

\setcounter{equation}{0}
\setcounter{section}{2}
\section*{Appendix \thesection. Computation of the Cohomology of $\BS$}

In Section 3, we met two cohomology problems: the first is~\reff{slacond} which
leads to the possible counterterms of the theory, and the second is~\reff{brsl}
which gives the anomaly of the theory. Both problems are to be solved in the
space of local integrated functional with canonical dimension four and
Faddeev-Popov charge $q$ (with $q=0$  or 1
according to which problem we are dealing with),
subject to the constraints~\reff{fant} and~\reff{wrcons}. We will denote these
constrained spaces by $\FF^{(q)}$.

The first method, which we applied to find $\S_c$ in~\reff{slacond},
consists in the following steps~:
\begin{itemize}
\item [1.]Write a basis $\{\Xi_i\}_{i\geq 1}$ of $\FF^{(-1)}$.
\item [2.]Apply $\BS$ on it to
      obtain the set of functionals $\{\BS\Xi_i\}_{i\geq 1}$,
\item [3.]Construct a basis $\{{\Theta}_i\}_{i\geq 1}$
          of the space spanned by
this set of functions. Then for each $i$ we can find a ${\hat \S}_i$ such that
\eq
\BS {\hat \S}_i = {\Theta}_i.
\end{equation}
The $\BS {\hat \S}_i$ are thus trivial solutions of the problem.
\item [4.]Complete the basis of step~3 by the set $\{\S^{\#}_i\}_{i\geq 1}$
such that this whole
set of functionals constitutes a basis of the invariant functionals of
$\FF^{(0)}$.
Since by construction there does not exist a $\Xi_i '$ such that
$\BS\Xi_i ' = \S^{\#}_i$, the functions
$\S^{\#}_i$ are non-trivial solutions of the problem.
\end{itemize}

The general solution of~\reff{slacond} can thus be written
\eq
\S_c = \dsum{i\geq 1}\a_i \BS {\hat \S}_i + \dsum{i\geq 1}\b_i \S^{\#}_i ,
\eqn{solsol}
where the $\a_i$, $\b_i$ are some constants.
This leads to the result \equ{contr} to \equ{diman}.

The advantage of this method is that it gives both the trivial and the
non-trivial part of the solution. The inconvenient is that it requires very
long
calculations  which can be partially avoided by the more sophisticated method
using filtration to which we now turn.

The second method, which we used to find $\D$ in~\reff{brsl}, was
developed in ~\cite{diximp}
and applied in~\cite{white2,magg1,magg2,white1,mpr}, and consists
first into passing
from functionals to functions, next to use a filtration to get a
simpler problem of local cohomology for the lowest order $\BSn{0}$ of the
operator $\BS$, and finally to
recover the cohomology of $\BS$.

This corresponds in practice into
translating the functional operator $\BS$, which acts on the
space of local functionals,
into an ordinary differential
operator, also denoted by $\BS$,
which acts on the space of functions $\D(x)$.
Thus, \reff{brsl} becomes a problem of local cohomology  modulo $d$,
\eq
\BS \D(x) + d\TD(x)=0,
\eqn{simo}
where $d$ is the exterior derivative: $d^2=0$, $\D(x)$ is
a 4-form defined by $\D=\int \D(x)$ and $\TD(x)$ is some 3-form with dimension
three and
$\F\Pi$ charge two.

\noindent{\bf Definition:} {\it
$\D$ is \Bmodd equivalent to $\D '$ if and only if
\eq
\D -\D ' =\BS\HD + d\HD ',
\end{equation}
for some $\HD$ and $\HD '$.
Moreover, $\D$ is called \Bmodd trivial if and only if
it is equivalent to zero.
}

\noindent{\bf Definition:} {\it
$\D$ is \Bmodd invariant if and only if there exists $\D '$ such that
\eq
\BS \D + d \D ' = 0 .
\end{equation}
}

\noindent{\bf Definition:} {\it
The cohomology of \Bmodd is then defined as the space of equivalence classes
of non-trivial invariant functions.
}

\noindent{\bf Remark:}{\em
Where no confusion may arise, we will speak indifferently of a function $\D$ or
of the class to which it belongs.}

The general solution of~\reff{simo} can  formally be written
\eq
\D(x) = \BS\HD(x) + d\HD '(x) + \CD(x) ,
\eqn{gensol}
where $\HD(x)$ and $\HD '(x)$ are some
forms with appropriate degree and $\F\Pi$ charge, and
$\CD$ belongs
to the cohomology of \Bmodd.

\noindent{\bf Definition:} {\it
Let $\NN$ be a ``filtration" operator mapping the spaces $\FF^q_p$ of
$p$-forms with $\F\Pi$ charge $q$ into itself, the
eigenvalues $n$ of $\NN$ being the nonnegative integers\footnote{
We restrict ourselves to the case of a filtration which commutes with $d$.}.
The expansion of a function $\D^q_p\in\FF^q_p$ and
of the operator $\BS$ according to
their eigenvalues reads
\eq
\ba{cc}
\BS = \dis{\sum_{n=0}^N} \BSn{n},&
\D^q_p =
\dsum{n\geq M} \D^{(n)} {}^q_p,
\ea
\end{equation}
with
\eq
\ba{cc}
\lc\NN,\BSn{n}\rc = n \BSn{n},
&
\NN \D^{(n)} {}^q_p = n  \D^{(n)} {}^q_p,
\ea
\eqn{filtn}
where $M$ stands for the order of the first non-vanishing term of the
expansion of $\D^q_p$.
}

By means of these definitions, we can filter~\reff{simo} into the following
set of equations
\eq\ba{c}
\BSn{0}\D^{(M)} {}^1_4 + d\D^{(M)} {}^2_3 = 0 ,\es
\BSn{1}\D^{(M)} {}^1_4 + \BSn{0}\D^{(M+1)} {}^1_4 + d\D^{(M+1)} {}^2_3 = 0 ,\es
\BSn{2}\D^{(M)} {}^1_4 + \BSn{1}\D^{(M+1)} {}^1_4 +
\BSn{0}\D^{(M+2)} {}^1_4 + d\D^{(M+2)} {}^2_3 = 0 ,\es
\mbox{\etc}
\ea\end{equation}
which can be written
at any order\footnote{In this context, the order of
an equation is the eigenvalue obtained by application of $\NN$ on it.}
$P$ (with $P\geq M$)
\eq
\ddsum{k=0}{\min \{N,P-M\}} \BSn{k} \D^{(P-k)} {}^1_4 +d\D^{(P)} {}^2_3 =0.
\eqn{ordrep}
The nilpotency of $\BS$~\reff{niloffsh} can be filtered into
\eq\ba{c}
\BSn{0}\BSn{0} =0 ,\es
\BSn{1}\BSn{0} + \BSn{0}\BSn{1} =0 ,\es
\BSn{2}\BSn{0} + \BSn{1}\BSn{1} + \BSn{0}\BSn{2}=0 ,\es
\vdots
\ea\eqn{b2}
Notice that $\BSn{0}$ is nilpotent,
as expressed by the first equation in~\reff{b2}, and
thus we can as well define the notions of \B0modd equivalence, invariance and
cohomology.

The procedure we will apply
to find the explicit form of $\CD(x)$ in~\reff{gensol}
consists in the following steps:

\noindent{\bf Step 1:}
It has been shown in~\cite{diximp} that for each element of the
cohomology of \Bmodd, we can pick up a particular representative which has the
property that its lowest order according to the filtration $\NN$
is \B0modd non-trivial. Thus, the first step
of our procedure is to
find the cohomology of \B0modd in the sector of 4-forms with $\F\Pi$ charge
one,
which is denoted by ${\widetilde{\FF}}_0$.
To this end, we apply the method described
at the beginning of this appendix for the
search of the counterterms but
now in the sector of ghost one and for the simpler
operator $\BSn{0}$. That means that
we begin by writing a basis ${\{\D_i{}^0_4\}}_{i\geq 1}$ of the space of
4-forms with $\F\Pi$ charge zero and apply
$\BSn{0}$  to these elements, thus getting
a set of functions ${\{\BSn{0}\D_i{}^0_4\}}_{i\geq 1}$. This set of functions
generate the space of \B0modd trivial invariant 4-forms
with $\F\Pi$ charge one.
We can then choose a basis of this
space and complete it in a basis of the space of
the \B0modd invariant 4-forms
with $\F\Pi$ charge one. The functions we just have defined to complete
the basis are non-trivial invariant 4-forms
with $\F\Pi$ charge one and therefore define a basis
of ${\widetilde{\FF}}_0$.

\noindent{\bf Remark:} {\em
Notice that as $\BSn{0}$ and $d$ both commute with $\NN$, the elements of
${\widetilde{\FF}}_0$ have a definite order and we will therefore denote them
by
$\CD^{(n)}$.}

\noindent{\bf Step 2:}
We then  proceed to the ``extension" of ${\widetilde{\FF}}_0$, and to this
aim we introduce the notion of the extension of an element
$\CD^{(n)}\in {\widetilde{\FF}}_0$, which consists
in constructing, if possible, the
$\BD^{(m)}$ for $m>n$ such that
\eq
\BD\equiv\CD^{(n)}+\dsum{m>n}\BD^{(m)}
\end{equation}
is \Bmodd invariant.
${\widetilde{\FF}}'$ is then defined as the space
generated by all the \Bmodd invariant functions with a \B0modd non-trivial
lowest order, i.e. by the extension of the elements of ${\widetilde{\FF}}_0$.
The explicit procedure of extension consists in solving~\reff{ordrep}
order by order in such a way that we get explicitly the elements
of
${\widetilde{\FF}}'$.

We proceed by induction: suppose we know
the general solutions $\D_{(P-1)}^{(\Pp)} {}^q_p$
satisfying~\reff{ordrep} till order $P-1$:
\eq\lac\ba{l}
\D_{(P-1)}^{(\Pp)} {}^1_4 = \BD_{(P-1)}^{(\Pp)} {}^1_4 +
\ddsum{k=0}{\min \{N,\Pp-M\}}\BSn{k} \HD^{(\Pp -k)} {}^0_4 +
d\HD^{(\Pp)}{ }^1_3 ,\es
\D_{(P-1)}^{(\Pp)} {}^2_3 = \BD_{(P-1)}^{(\Pp)} {}^2_3 +
\ddsum{k=0}{\min \{N,\Pp-M\}}\BSn{k} \HD^{(\Pp -k)} {}^1_3 +
d\HD^{(\Pp)}{ }^2_2 ,
\ea\right .\quad\mbox{for}\; M\leq \Pp\leq P-1,
\eqn{solordrepp}
where the $\HD^{(\Pp -k)} {}^q_p$ are some functions for which we do not add a
lower index since, as we will see below,
they get neither constraints nor corrections at the next order.
The $\BD_{(P-1)}^{(\Pp)} {}^q_p$ constitute thus the beginning of the
extension we are looking for.
Inserting these solutions in~\reff{ordrep} at order $P$
and using the nilpotency of $\BS$ expressed by~\reff{b2} gives
\eqap
\BSn{0} \D^{(P)} {}^1_4
+ d \D^{(P)} {}^2_3
+ \ddsum{k=1}{\min \{N,P-M\}} \BSn{k} \BD_{(P-1)}^{(P-k)} {}^1_4 &&\nonumber\es
- \BSn{0}\lp\ddsum{k=1}{\min \{N,P-M\}} \BSn{k} \HD^{(P-k)} {}^0_4
\rp
- d\lp\ddsum{k=1}{\min \{N,P-M\}} \BSn{k} \HD^{(P-k)} {}^1_3
\rp&=&0 ,
\label{constraint}\eea
which entails the following interpretation:
first it imposes constraints on the free coefficients of the general
solution~\reff{solordrepp} since
the third term of~\reff{constraint} has to be \B0modd trivial, and secondly we
see that the terms involving $\HD$ do not suffer any constraint to be extended
at order $P$ since they appear in~\reff{constraint} in a trivial form.
Let us denote by $\BD_{(P)}^{(\Pp)} {}^1_4$
the constrained lower order solutions
and by $\TD^{(P)} {}^1_4$,
$\TD^{(P)} {}^2_3$ a special solution
of
\eq
\ddsum{k=1}{\min \{N,P-M\}} \BSn{k} \BD_{(P)}^{(P-k)} {}^1_4 =
- \BSn{0} \TD^{(P)} {}^1_4
- d \TD^{(P)} {}^2_3 .
\end{equation}
The general solution of~\reff{constraint} then reads
\eq\lac\ba{l}
\D_{(P)}^{(P)} {}^1_4 =
\TD^{(P)} {}^1_4 +
\CD^{(P)} {}^1_4 +
\ddsum{k=1}{\min \{N,P-M\}} \BSn{k} \HD^{(P-k)} {}^0_4 +
\BSn{0} \HD^{(P)} {}^0_4 +
d\HD^{(P)}{ }^1_3 ,\es
\D_{(P)}^{(P)} {}^2_3 =
\TD^{(P)} {}^2_3 +
{\D '}^{(P)} {}^2_3 +
\ddsum{k=1}{\min \{N,P-M\}} \BSn{k} \HD^{(P-k)} {}^1_3 +
\BSn{0} \HD^{(P)} {}^1_3 +
d\HD^{(P)}{ }^2_2 ,
\ea\right. \eqn{genp2}
where $\CD^{(P)} {}^1_4$ belongs to
${\widetilde{\FF}}_0$, and ${\D '}^{(P)} {}^2_3$
satisfies
\eq
\BSn{0} \CD^{(P)} {}^1_4 + d{\D '}^{(P)} {}^2_3=0.
\end{equation}
Defining $\BD_{(P)}^{(P)} {}^q_p\equiv\TD^{(P)} {}^q_p +
\CD^{(P)} {}^q_p$ leads us to the general
solution of~\reff{ordrep} at order $P$
in the same form as in~\reff{solordrepp}, thus ending the induction step.

\noindent{\bf Remark:} {\em
In principle this system of equations is endless but in practice it
becomes rapidly trivial due to dimensional constraints (all the
functions under consideration have dimension four), the spaces
${\FF}^{(n)}$ of functions of order $n$ being empty for $n$ greater than some
constant (in our case it will be five).}

Then we sum the solutions $\D^{(P)}_{(\infty)}$ given by~\reff{solordrepp}
to get the general solution\\
$\D=\dsum{P\geq M}\D^{(P)}_{(\infty)}$, which we can write in the form
\eq
\D = \BD + \dsum{P\geq M}\lp\BS\HD^{(P)}{}^0_4+d\HD^{(P)}{}^1_3\rp,
\end{equation}
which shows that the extensions of the \B0modd trivial terms
$\BSn{0}\HD^{(P)}{}^0_4+d\HD^{(P)}{}^1_3$ are the \Bmodd trivial terms
$\BS\HD^{(P)}{}^0_4+d\HD^{(P)}{}^1_3$ and that the general element
of ${\widetilde{\FF}} '$ we are looking for is $\BD$.

We then define the subspaces ${\widetilde{\FF}}'_{(n)}$ of ${\widetilde{\FF}}'$
as formed by the elements whose lowest order is~$n$.

\noindent{\bf Step 3:}
For each element $\BD_{(n)}\in{\widetilde{\FF}}'_{(n)}$, which is invariant
by construction, we have to separate the
\Bmodd non-trivial part from the trivial one.
To this aim, let us introduce the notion of triviality {\em until order $M$}:

\noindent{\bf Definition:} {\it
$\D^{(n)}$ is \Bmodd trivial until order $n$ if and only if
\eq
\exists\;\HD ,\;\HD '\makebox[3cm]{such that}
\D^{(n)} = \BS\HD + d\HD ' + O(n + 1) ,
\eqn{crittriv}
where $O(n + 1)$ contains terms of order greater than $n$.
}

It leads to the following decomposition
\eq
\BD_{(n)} = \BD_{(n)}' + \BD_{(n)}^\#,
\eqn{deco}
where $\BD_{(n)}'$, contrarily to $\BD_{(n)}^\#$, has a lowest
non-vanishing order $\BD_{(n)}'{}^{(n)}$
which is \Bmodd trivial till order~$n$. It follows that~$\BD_{(n)}'$
is \Bmodd trivial and that~$\BD_{(n)}^\#$ constitutes the anomaly.
Finally, we can write the
explicit form of $\CD$ in~\reff{gensol} as~:
\eq
\CD = \dsum{n\geq M} \b_{(n)}\BD_{(n)}^\#\ ,
\end{equation}
where the $\b_{(n)}$ are some arbitrary constants.

We have now finished with the description of the general method and
turn to the specific case of interest
in this paper.
We choose a filtration
operator $\NN$ which assigns the weight 1 to all the fields and their
derivatives as
well as to the global ghosts $\e^\a$, $\eb^\db$, $\xi^\m$ and $\eta$~:
\eq
\NN = \int\dsum{\mbox{\scriptsize{all fields}}\; \f}\lac\f\dpad{}{\f}
+\pam\f\dpad{}{\pam\f}+\cdots\rac
+ \e^\a\pad{}{\e^\a} +
\eb^\db\pad{}{\eb^\db} +
\xi^\m \pad{}{\xi^\m} +
\eta \pad{}{\eta}.
\end{equation}
The $\BSn{0}$-transformations of the
fields are (the transformations of their derivatives
being trivially inferred due to this simple choice of the weights)
\eq\ba{lccclcll}
\BSn{0}A^{i\m}&=&\partial^\m c^i,&\makebox[10mm]{}&
\BSn{0} {\hat A}^{*}{}^{i\m}&=&&
\frac{1}{g^2}\pan\lp\partial^\n A^{i\m} -\partial^\m A^{i\n}\rp,\es
\BSn{0}c^i&=&0,&&\BSn{0}c^{*i}&=&&\pam A^{*}{}^{i\m},\es
\BSn{0}\l^{i\a}&=&0,&&\BSn{0}\lb^{*}{}^i_\db&=&-&
        \frac{i}{g^2}\pam\l^{i\a}\smab,\es
\BSn{0}\lb^{i\db}&=&0,&&\BSn{0}\l^{*}{}^i_\a&=&
       &\frac{i}{g^2}\smab\pam\lb^{i\db},\es
\BSn{0}\f_a&=&0,&&\BSn{0}\fb^{*}_a&=&-&\frac{1}{2}\pam\partial^\m\f_a,\es
\BSn{0}\fb_a&=&0,&&\BSn{0}\f^{*}_a&=&-&\frac{1}{2}\pam\partial^\m\fb_a,\es
\BSn{0}\p^\a_a&=&0,&&\BSn{0}\Nb_{a \db}&=&-&i\pam\p^\a_a\smab,\es
\BSn{0}\pb^\db_a&=&0,&&\BSn{0}\p^{*}{}_{a \a}&=&&i\smab\pam\pb^\db_a,\es
\BSn{0}b^i&=&0,&&\BSn{0}\cb^i&=&&0,\es
\BSn{0}\e^\a&=&0,&&\BSn{0}\xi^\m&=&&0,\es
\BSn{0}\eb^\db&=&0,&&\BSn{0}\eta&=&&0.\es
\ea\eqn{b0var}
Now, beginning with Step~1, we find the general elements
of the cohomology of \B0modd at each order:
\eqap
\D^{\#(3)} {}^1_4&=&\a_1d^{ijk}\mnrs\pam c^i A^j_\n \partial_\r A^k_\s,\es
\D^{\#(4)} {}^1_4&=&\a_2d^{ijk}\lp\e^\a\l^i_\a\rp\lp\lb^j_\db\lb^{k\db}\rp+
\a_3d^{ijk}\lp\eb_\db\lb^{i\db}\rp\lp\l^{j\a}\l^k_\a\rp\nonumber\es
&+&\a_{4a(bc)}\lp\e^\a\p_{a \a}\rp\lp\p^\b_b\p_{c \b}\rp+
\a_{5a(bc)}\lp\eb_\da\pb^\da_a\rp\lp\pb_{b \db}\pb^\db_c\rp\nonumber\es
&+&\a^i_{6ab}\lp\e^\a\p_{a \a}\rp\lp\lb^i_\db\pb^\db_b\rp+
\a^i_{7ab}\lp\eb_\db\pb^\db_a\rp\lp\l^{i\a}\p_{b \a}\rp\nonumber\es
&+&\a^i_{8ab}\lp\e^\a\smab\lb^{i \db}\rp\f_a\pam\fb_b+
\a^i_{9ab}\lp\l^{i \a}\smab\eb^\db\rp\fb_a\pam\f_b\nonumber\es
&+&\a^i_{10ab}\lp\eb^\da{\bar{\s}}^{\m\n}_{\da\db}\pb^\db_b\rp\f_a\pam A^i_\n+
\a^i_{11ab}\lp\e^\a\s^{\m\n}_{\a\b}\p^\b_b\rp\fb_a\pam A^i_\n\nonumber\es
&+&\a_{12a[bc]}\lp\e^\a\smab\pb^\db_a\rp\pam\fb_b\fb_c+
\a_{13a[bc]}\lp\p^\a_a\smab\eb^\db\rp\pam\f_b\f_c\nonumber\es
&+&\lp\esmab\rp\lac\a_{14}\lp c^{*i}\pam c^i-
{\hat A}^{*}{}^i_\n\pam A^{i\n}\rp+
\a_{15}\lp\l^{*}{}^{i\a}\pam\l^i_\a+\lb^{*}{}^i_\db\pam\lb^{i\db}\rp\right.
\nonumber\es
&+&\left.\a_{16}\lp -\f^{*}_a\pam\f_a-\fb^{*}_a\pam\fb_a\rp+
\a_{17}\lp\p^{*}{}^\a_a\pam\p_{a\a}+\Nb_{a\db}\pam\pb^\db_a\rp\rac,\es
\D^{\#(5)} {}^1_4&=&\a^i_{18(abc)}\lp\e^\a\l^i_\a\rp\fb_a\fb_b\fb_c+
\a^i_{19(abc)}\lp\eb_\da\lb^{i\da}\rp\f_a\f_b\f_c\nonumber\es
&+&\a_{20ab(cd)}\lp\e^\a\p_{a\a}\rp\f_b\fb_c\fb_d+
\a_{21(ab)cd}\f_a\f_b\lp\eb_\da\pb^\da_c\rp\fb_d,
\eea
where  $d^{ijk}$  is defined by~\reff{dijk}.
Doing the extension of these functions\footnote{
To do this extension we need of course the explicit form of $\BSn{n}$ for
$n\geq 1$, which can be read out from~\reff{brs}, \reff{classact}
and~\reff{slavnovlin}
(in our case $\BS=\BSn{0}+\BSn{1}+\BSn{2}$).} according to Step~2
gives rise to
the following constraints on the coefficients $\a_n$:
\eq
\a_1=-\frac{1}{3}\a_2=\frac{1}{3}\a_3\equiv r,
\end{equation}
\eq
\a_{4a(bc)}= \a_{4(abc)},
\eqn{a4}
\eq
\a_{5a(bc)}= \a_{5(abc)},
\eqn{a5}
\eq
\makebox[3mm]{}\frac{1}{2}\a^i_{6ab}=
    \makebox[3mm]{} i \a^i_{8ab}=\a^i_{10ab}\equiv \a_{\rm I}{}^i_{ab},
\end{equation}
\eq
-\frac{1}{2}\a^i_{7ab}= - i \a^i_{9ab}=\a^i_{11ab}\equiv \a_{\rm II}{}^i_{ab},
\end{equation}
\eq
\a_{12}=\a_{13}=\a_{18}=\a_{19}=0,
\end{equation}
\eq
\a_{14}=\a_{15}=\a_{16}=\a_{17}\equiv  \a_{\rm III},
\end{equation}
\eq
\a_{20ab(cd)}= - \a_{21(ab)cd}\equiv \a_{\rm IV}{}_{(ab)(cd)},
\end{equation}
where $\a_{\rm I}{}^i_{ab}$, $\a_{\rm II}{}^i_{ab}$ and
$\a_{\rm IV}{}_{(ab)(cd)}$ are invariant tensors and $\a_{\rm III}$ an
arbitrary
constant.
(From~\reff{a4} and~\reff{a5} follows
that the terms in $\a_4$ and $\a_5$ vanish).
This shows that the space
${\widetilde{\FF}}'$ defined in Step~2 is of dimension~5.

We choose for ${\widetilde{\FF}}'$ the basis elements $\D_{\mbox{\tiny{SAB}}}$,
$\BD_{\rm I}, \cdots , \BD_{\rm IV}$ characterized by their lowest order terms
$\D^{\#}_{\mbox{\tiny{SAB}}}{}^{(3)}$,
$\D^{\#}_{\rm I}{}^{(4)}, \cdots, \D^{\#}_{\rm IV}{}^{(5)}$ given by:
\eqap
\D^{\#}_{\mbox{\tiny{SAB}}}{}^{(3)}&=&d^{ijk}\mnrs\pam c^i A^j_\n
\partial_\r A^k_\s,\es
\D^{\#}_{\rm I}{}^{(4)}&=&\a_{\rm I}{}^i_{ab}
\lp 2 \e^\a\p_{a \a}\lb^i_\db\pb^\db_b
-i \e^\a\smab\lb^{i \db}\f_a\pam\fb_b
+ \eb^\da{\bar{\s}}^{\m\n}_{\da\db}\pb^\db_b\f_a\pam A^i_\n
\rp,\es
\D^{\#}_{\rm II}{}^{(4)}&=&\a_{\rm II}{}^i_{ab}
\lp -2 \eb_\db\pb^\db_a\l^{i\a}\p_{b \a}
+i \l^{i \a}\smab\eb^\db\fb_a\pam\f_b
+ \e^\a\s^{\m\n}_{\a\b}\p^\b_b\fb_a\pam A^i_\n
\rp,\es
\D^{\#}_{\rm III}{}^{(4)}&=&-i\a_{\rm III}
\lp\esmab\rp
\lp
- {\hat A}^{*}{}^i_\n\pam A^{i\n}
+ \l^{*}{}^{i\a}\pam\l^i_\a
+ \lb^{*}{}^i_\db\pam\lb^{i\db}
+ c^{*i}\pam c^i
\right.\nonumber\es
&&\makebox[26mm]{}\left.
-\f^{*}_a\pam\f_a
-\fb^{*}_a\pam\fb_a
+\p^{*}{}^\a_a\pam\p_{a\a}
+\Nb_{a\db}\pam\pb^\db_a
\rp,\es
\D^{\#}_{\rm IV}{}^{(5)}&=&\a_{\rm IV}{}_{(ab)(cd)}
\lp\lp\e^\a\p_{a\a}\rp\f_b\fb_c\fb_d-\f_a\f_b\lp\eb_\da\pb^\da_c\rp\fb_d
\rp.
\eea

Next, according to Step~3,
we test the triviality of these elements. We find that
$\D_{\mbox{\tiny{SAB}}}$ is \Bmodd non-trivial,
and that
\eqap
\D^{\#}_{\rm I}{}^{(4)}&=&\BS
\lp \a_{\rm I}{}^i_{ab} \f_a \lb^i_\db\pb^\db_b \rp
+ d \lp i\a_{\rm I}{}^i_{ab} \xi\f_a \lb^i_\db\pb^\db_b \rp
+ O(5),\es
\D^{\#}_{\rm II}{}^{(4)}&=&\BS
\lp \a_{\rm II}{}^i_{ab} \fb_a \l^{i\a}\p_{b \a} \rp
+ d \lp i\a_{\rm II}{}^i_{ab} \xi \fb_a \l^{i\a}\p_{b \a} \rp
+ O(5),\es
\D^{\#}_{\rm III}{}^{(4)}&=&\BD_{\rm III}=
\BS\lp \a_{\rm III}\LL\rp + d\lp i\a_{\rm III}\xi\LL\rp ,\es
\D^{\#}_{\rm IV}{}^{(5)}&=&\BD_{\rm IV}=\BS
\lp \frac{1}{4}\a_{\rm IV}{}_{(ab)(cd)}\f_a\f_b\fb_c\fb_d \rp
+ d \lp \frac{i}{4}\a_{\rm IV}{}_{(ab)(cd)}\xi\f_a\f_b\fb_c\fb_d \rp
,
\eea
where $\LL $ is defined up to a total derivative by
\eq
\intx \LL = \frac{1}{4}\lp\S-\intx b^i\pam A^{i\m} -\eta\D_{\rm R}
-\xi^\m\D^{\rm t}_{\m}\rp.
\end{equation}
This entails the \Bmodd triviality of the four elements
$\BD_{\rm I}, \cdots , \BD_{\rm IV}$, and therefore $\D_{\mbox{\tiny{SAB}}}$
constitutes a basis of the cohomology of \Bmodd.

We finally write the explicit form of the anomaly:
\eqa
\D_{\mbox{\tiny{SAB}}}= \LP&\!\!\!
\mnrs\lac d^{ijk}\lp\pam c^i -
2\l^{i\a}\s_{\m\a\db}\eb^\db-2\e^\a\s_{\m\a\db}\lb^{i\db}\rp
A^j_\n \partial_\r A^k_\s\right.\nonumber\es &\!\!\!
+\left.D^{ijmk}\lp\frac{1}{12}\pam c^i-\frac{1}{4}
\l^{i\a}\s_{\m\a\db}\eb^\db-\frac{1}{4}\e^\a\s_{\m\a\db}\lb^{i\db}\rp A^j_\n
A^m_\r A^k_\s\rac\nonumber\es &\!\!\!
-3d^{ijk}\lp\e^\a\l^i_\a\lb^j_\db\lb^{k\db}-
\eb_\db\lb^{i\db}\l^{j\a}\l^k_\a\rp\RP,
\eqan{sabloc}
where  $d^{ijk}$  and  $D^{ijmk}$  are defined by~\reff{dijk} and~\reff{Dijkm}
respectively.

\setcounter{equation}{0}
\setcounter{section}{3}
\section*{Appendix \thesection . Notations and
                   Conventions}\label{notations}
\newcommand{\point}[1]{\vspace{3mm}

\noindent{\bf #1}}
\newcommand{\iv}{\dint dV\;}
\newcommand{\is}{\dint dS\;} \newcommand{\isb}{\dint d\bar S\;}
\newcommand{\ad}{{\dot\a}}  \newcommand{\bd}{{\dot\b}}
\newcommand{\gd}{{\dot\g}}
\newcommand{\ed}{{\dot\e}}
\newcommand{\QB}{{\bar{Q}}}  \newcommand{\WB}{{\bar{W}}}
\newcommand{\DB}{{\bar{D}}}
\newcommand{\BBAR}{{\bar{B}}}
\newcommand{\SB}{{\bar{S}}}
\newcommand{\AB}{{\bar{A}}}
\newcommand{\JB}{{\bar{J}}}
\renewcommand{\sb}{{\bar\s}}

\newcommand{\tb}{{\bar\th}}
\renewcommand{\pb}{{\bar\p}} \newcommand{\chib}{{\bar\chi}}
\newcommand{\smuaad}{\s^\m_{\a\ad}}
\newcommand{\sbmuaad}{{\bar\s}_\m^{\ad\a}}
\point{Units:} $\hbar=c=1$
\point{Space-time metric:} $(g_{\m\n}) = \mbox{diag}(1,-1,-1,-1)
    \ ,\quad(\m,\n,\cdots=0,1,2,3)$
\point{Fourier transform:}
\[
f(x) = \dfrac{1}{2\pi}\int dk\;e^{ikx}\tilde f(k)\ ,\quad
\tilde f(k) = \int dx\;e^{-ikx}f(k)\ .
\]
\point{Weyl spinor:} $(\p_\a\ ,\ \a =1,2)\
   \in\, \mbox{repr. }(\half,0)$ of the Lorentz group. \\
  The spinor components are Grassmann variables: $\p_\a\p'_\b=-\p'_\b\p_\a$
\point{Complex conjugate spinor:}
    $(\bar\p_\ad\ ,\ \ad =1,2)\
                           \in\, \mbox{repr. }(0,\half)$
\point{Raising and lowering of spinor indices:}
  \[\ba{l}
  \p^\a=\e^{\a\b}\p_\b\ ,\quad \p_\a=\e_{\a\b}\p^\b\ ,\es
  \mbox{with }   \e^{\a\b}=-\e^{\b\a}\ ,\quad\e^{12}=1\ ,\quad
  \e_{\a\b}=-\e^{a\b}\ ,\quad \e^{\a\b}\e_{\b\g}=\d^\a_\g\ ,\es
  \mbox{(the same for dotted indices).}
  \ea\]
\point{Derivative with respect to a spinor component:}
  \[\ba{l}
  \dpad{}{\p^\a}\p^\b=\d^{\b}_{\a}\ ,\quad
     \dpad{}{\p_\a} = \e^{\a\b}\dpad{}{\p^\b}\ ,\es
  \mbox{(the same for dotted indices)}
  \ea\]
\point{Pauli matrices:}
  \[\ba{l}
  \lp\s^\m_{\a\bd}\rp=
  \lp\, \s^0_{\a\bd},\, \s^1_{\a\bd},\,
              \s^2_{\a\bd},\, \s^3_{\a\bd}\, \rp \es
  \sb_\m^{\ad\b}=\s_\m^{\b\ad}
    =\e^{\b\a}\e^{\ad\bd}\s_{\m\,\a\bd}\ ,\es
   {(\s^{\m\n})_\a}^\b=
    \dfrac{i}{2}{\lc \s^\m\sb^\n-\s^\n\sb^\m \rc_\a}^\b   \ , \quad
  {(\sb^{\m\n})^\ad}_\bd=
    \dfrac{i}{2}{\lc \sb^\m\s^\n-\sb^\n\s^\m \rc^\ad}_\bd   \ ,
\ea\]
with
  \[\ba{l}
  \s^0=\lp\matrix{1&0\\0&1}\rp\ ,\quad  \s^1=\lp\matrix{0&1\\1&0}\rp\ ,\quad
    \s^2=\lp\matrix{0&-i\\i&0}\rp\ ,\quad
                \s^3=\lp\matrix{1&0\\0&-1}\rp\ ,\\[5mm]
  \sb^0 =\s^0\ ,\quad \sb^i=-\s^i=\s_i\ ,
  \s^{0i} = -\sb^{0i} = -i\s^i\ ,\quad
   \s^{ij} = \sb^{ij} = \e^{ijk}\s^k\ ,\es
   \quad i,j,k=1,2,3\ .
  \ea\]
\point{Summation conventions and complex conjugation:}
Let $\p$ and $\chi$ be two Weyl spinors.
  \[\ba{l}
  \p\chi=\p^\a\chi_\a = -\chi_\a\p^\a = \chi^\a\p_\a = \chi\p \ ,\es
  \pb\chib=\pb_\ad\chib^\ad = -\chib^\ad\pb_\ad
        = \chib_\ad\pb^\ad  = \chib\pb \ ,\es
  \p\s^\m\chib = \p^\a\smuaad\chib^\ad\ ,\quad
        \pb\sb_\m\chi = \pb_\ad\sbmuaad\chi_\a\ ,\es
  \overline{(\p\chi)} = \chib\pb = \pb\chib \ ,\es
  \overline{(\p\s^\m\chib)} = \chi\s^\m\pb = -\pb\sb^\m\chi\ ,\es
  \overline{(\p\s^{\m\n}\chi)} = \chib\sb^{\m\n}\pb\ .
\ea\]
\point{Infinitesimal Lorentz transformations of
         the Weyl spinors:}
  \[\ba{l}
  \mbox{if}\quad
  \d^{\rm L} x^\m = {\o^\mu}_\n x^\n\ ,\quad
     \mbox{with }\o^{\m\nu} = -\o^{\n\m}\qquad
      \lp\o^{\m\n}=g^{\n\r}{\o^\m}_\r\rp \ ,\quad\mbox{then:}
  \es
  \d^{\rm L}\p_\a(x) =
    \half\o^{\m\n}\lp (x_\m\pa_\n - x_\n\pa_\m)\p_\a(x)
      -\dfrac{i}{2}{(\s_{\m\n})_\a}^\b \p_\b \rp\ ,\es
  \d^{\rm L}\pb^\ad(x) =
    \half\o^{\m\n}\lp (x_\m\pa_\n - x_\n\pa_\m)\pb^\ad(x)
      +\dfrac{i}{2}{(\sb_{\m\n})^\ad}_\bd \pb^\bd \rp\ .

\ea\]
\setcounter{equation}{0}
\setcounter{section}{5}
\section*{ERRATUM}
\renewcommand{\lb}{{\bar\lambda}}
\renewcommand{\fb}{{\bar\phi}}
\renewcommand{\pb}{{\bar\psi}}
\newcommand{\susy}{supersymmetry}
\newcommand{\Susy}{Supersymmetry}
\newcommand{\susyc}{supersymmetric}
\newcommand{\Susyc}{Supersymmetric}
\hfill April 1996

In the above paper we came to an erroneous conclusion  concerning
the unphysical part of the counterterm (3.30). 
 We are in fact not able to prove algebraically that, in the
Wess-Zumino gauge, the anomalous dimensions of both partners of an 
$N=1$
supersymmetry multiplet are equal, as they are indeed in a 
superspace
formulation or more generally in a formulation where the
 supersymmetry
is realized linearly. 
That this equality may actually not hold~\cite{atv} 
in the Wess-Zumino
gauge has not to be considered as a contradiction with 
supersymmetry
since the anomalous dimensions are in general gauge dependent
quantities.

The source of the mistake is the wrong result given in Appendix A. 
This result, as it was expressed by the   equations (A.3) and
(A.4) of this appendix, was that the linear part of the \susy\
transformations, namely the \susy\ transformations of the bosonic 
fields,
was not renormalized. The argument presented, using a diagrammatic
analysis, was based on the ``observation'' that the 
terms 
\eq
(A_\m^{*i}+\pa_\m\cb)  \e\s^\m\lb^i\ ,\quad 2\f^*\e\p\ ,
\eqn{class-vertices}
in the classical action which describe these transformations being
linear in the quantum fields -- the other factors being external 
fields
or parameters --
cannot enter in any one-particule irreducible graph.
Here  $A_\m^*$ and $\f^*$ are  the external field coupled to the 
BRS transformations -- which include the \susy\ transformations -- of 
the gauge field $A_\m$ and of the scalar matter field $\f$,
$\l_\a$ is the gaugino field, $\p_\a$  the matter spinor field,
$\cb$ the antighost and
$\e_\a$ the infinitesimal parameter of the \susy\ transformations.

However the first of these vertices is
not linear in the quantum fields due to the external 
field $A^*$ being
shifted by the gradient of the antighost field $\cb$, which 
propagates!
The true result is that both interactions \equ{class-vertices} do 
get
radiative corrections.

 The consequence of this fact is that the counterterms 
in Eq. (3.30) corresponding to the renormalizations of the
boson fields and of the fermion fields, respectively, 
actually appear with independent coefficients. This does not modify our
conclusions concerning the renormalizability of the theory, the only
change being 
that the correct expression for the Callan-Symanzik equation
(Eq. (6.4)) must read
\eq\ba{c}
\Lp \m\dpad{}{\mu} + \b_g\dpad{}{g} + \b_{abc}\dpad{}{\l_{abc}}
+ \bar\b_{abc}\dpad{}{\lb_{abc}} 
- \g^A\NN^A - \g^\l\NN^\l  - \g^\lb\NN^\lb \es
- \g^\f_{ab}\NN^\f_{ab} - \g^\p_{ab}\NN^\p_{ab}
- \g^\fb_{ab}\NN^\fb_{ab} - \g^\pb_{ab}\NN^\pb_{ab}
\Rp\G=0\ ,
\ea\eqn{Callan-Sym-corr}
where
\[\ba{l}
\NN^\vf = \intx\lp\vf\dfud{}{\vf} - \vf^*\dfud{}{\vf^*}\rp \ ,
   \quad \vf=A,\l,\lb\ ,\es
\NN^\vf_{ab} = \intx\lp \vf_a\dfud{}{\vf_b} - 
\vf^*_b\dfud{}{\vf^*_a}\rp\ ,
   \quad \vf =\f,\p,\fb,\pb\ ,
\ea\]
the indices $a$, $b$ being group representation indices.



\end{document}